\documentclass[12pt]{article}
\usepackage{a4wide,feynarts,amsmath,amssymb,graphicx,psfrag}

\newcommand{\RE}{\text{Re}}

\newcommand{\TE}[1]{\cdot 10^{#1}}
\newcommand{\gev}{\text{GeV}}
\newcommand{\tev}{\text{TeV}}

\newcommand{\MIN}{{\text{min}}}

\newcommand{\JET}{\text{jet}}

\begin{document}
\setlength{\unitlength}{1mm}

\begin{titlepage}
\begin{flushright}
FR-PHENO-2010-016
\end{flushright}
\vspace{1cm}

\begin{center}
{\Large \bf
Electroweak and Bottom Quark Contributions\\
to Higgs Boson plus Jet Production \\
}
\vspace{2.5cm}
{\large \sc Oliver~Brein 
}\\[0.8cm]
{\normalsize\em 
Physikalisches Institut,
Albert-Ludwigs-Universit\"at Freiburg\\
Hermann-Herder-Str. 3, D-79104 Freiburg im Breisgau, Germany\\[2mm]
{\tt Oliver.Brein@physik.uni-freiburg.de}}\\[1cm]
\end{center}

\vspace{1cm}

\begin{abstract}
\noindent
This paper presents predictions for 
jet pseudorapidity ($\eta$) and transverse momentum ($p_T$) 
distributions 
for the production of the Standard Model
Higgs boson in association with a high-$p_T$ hadronic jet.
We discuss the contributions of electroweak loops
and of 
bottom-quark parton
processes to the cross section. The latter arise in the five-flavour scheme.
Predictions for the Tevatron and the Large Hadron Collider 
with $10\,\tev$ collision energy are presented.
For Higgs boson masses of $120\,\gev$, $160\,\gev$ and $200\,\gev$, 
we find the maximal effects 
of the electroweak contributions to the 
Higgs plus jet $p_T$ and $\eta$ distribution to be 
$-14\%$ and $-5.3\%$, respectively, for the Tevatron, 
and
$-3\%$ and $-2\%$, respectively, for the LHC.
For the maximal contribution of bottom-quark parton processes
to the $p_T$ and $\eta$ distribution, we find
$+3\%$ and $+2.5\%$, respectively, for the Tevatron,
and
$+3.5\%$ and $+3\%$, respectively, for the LHC.
A separate study of the Higgs + $b$-jet cross section
demonstrates that a calculational approach which respects the 
hierarchies of Yukawa couplings yields a leading order
cross section prediction which is more accurate 
in the high-$p_T$ regime than conventional approaches.
\end{abstract}

\end{titlepage}

\section{Introduction}

Since the discovery of the massive electroweak gauge bosons $Z$ and $W^\pm$
\cite{WZ-discovery},
we are sure to observe electroweak symmetry breaking in nature.
However, the underlying dynamics of this symmetry breaking has not yet been
established experimentally.
Spontaneous symmetry breaking via the Higgs mechanism \cite{higgs-mechanism} 
is an appealing theoretical suggestion for such a dynamics.
Depending on the specific model considered, 
the Higgs mechanism implies the existence of 
one or more scalar particles (Higgs bosons).
Therefore, the search for Higgs bosons is an important task
in the quest to unravel the nature of electroweak symmetry breaking
at high energy collision experiments.

Besides inclusive single Higgs production, 
Higgs boson production in association with a high-$p_T$ hadronic jet
provides a useful channel for Higgs searches at hadron colliders,
which allows for refined cuts increasing the signal-to-background ratio.
Specifically, for the Standard Model (SM) Higgs plus jet 
process~\cite{SMHJcalc,BaurGlover}, 
early simulations considering the decay channels
$H\rightarrow \gamma\gamma$ \cite{ADIKSS,Zmushko} 
and $H\rightarrow \tau^+ \tau^-$ \cite{mellado-etal}
have demonstrated promising signal-to-background ratios.
While the promising findings of the parton-level study \cite{ADIKSS} 
were confirmed by a simulation using the fast (and less detailed) 
simulation for the ATLAS detector response \cite{Zmushko}, 
an up-to-date simulation of the detailed ATLAS detector performance
appeared in \cite{ATLAS-CSC-book}.
From the CMS collaboration, only a simulation of the 
inclusive $H\rightarrow \gamma\gamma$ channel is currently available
\cite{CMS-TDR}.

Considerable progress has been made 
in improving the SM cross section predictions:
The fully differential distribution for Higgs production
is available
at next-to-next-to-leading order (NNLO) QCD accuracy in the 
large top-mass limit \cite{AMP}, improved by the 
resummation of logarithmically enhanced  
terms for low $p_T$~\cite{grazzini-etal}.
Dedicated calculations of 
higher-order corrections to differential cross sections for 
Higgs boson production associated with a high-$p_T$ jet
have been performed: the next-to-leading order (NLO) QCD corrections 
in the large top-mass limit \cite{SMHJatNLO}
and the corresponding resummation of 
large threshold logarithms
\cite{kulesza-etal}.
Furthermore, the soft-plus-virtual gluon approximation 
to the NLO QCD prediction of the Higgs $p_T$ distribution
\cite{SMHJatNLO-SplusV-approx} goes beyond the large top-mass limit.
For the $b$-quark process $b g\to H b$, 
the NLO QCD corrections 
are also known \cite{bghb-QCD1,bghb-QCD2},
and very recently a calculation of the weak corrections to this process
appeared \cite{Dawson-Jaiswal}.

The remaining scale uncertainty of the NLO QCD result
for the Higgs plus high-$p_T$ jet production cross section
via the light quark parton processes
is at the level of 10\% \cite{SMHJatNLO,kulesza-etal}.
The achievement of this level of accuracy in the QCD corrections,
albeit in the large top-mass limit,
motivates to study other 
leading order contributions which could potentially
affect the Higgs plus jet cross section prediction 
at the 10\% level.
Specifically, there are to consider: 
loop-induced electroweak contributions to 
the processes $qg\to Hq$ and $q\bar q\to H g$ 
for the light quark flavours ($u, d, s, c$)
and the contributions
from the processes $bg\to Hb$ and $b\bar b\to H g$. 
All these contributions may change the differential distributions 
of the process, which are therefore of particular interest.
Parts of these contributions to the Higgs plus jet cross section
have been considered elsewhere.
Electroweak and non-zero quark-mass effects on the Higgs $p_T$ distribution
using 5-flavour parton distribution functions (PDFs)
neglecting the contributions from $b$-quark parton processes
have been studied for the LHC and the Tevatron in \cite{keung-petriello}.
Leading order QCD and electroweak contributions to the 
$qg$ and $q\bar q$ processes,
including 
electroweak contributions
to the $b$-quark processes
enhanced by the top-Higgs Yukawa coupling, 
have been studied for the LHC 
in \cite{Mrenna-Yuan}. However, the impact of those contributions 
on the total Higgs plus jet cross section, which involves also the 
dominant gluon fusion channel, has not been studied in \cite{Mrenna-Yuan}.
To our knowledge, a combination of the contributions of 
the light quark flavours ($u,d,s,c$) and the bottom quarks
in order to get a cross section prediction for 
Higgs plus (any) jet in a 5-flavour scheme has 
so far only been done for 
the leading-order QCD predictions \cite{hjet-own,hjet-own-distributions}.

In the present paper, 
we study the combined effect of contributions of leading order (LO) 
electroweak loop graphs and bottom-quark parton processes on the 
jet pseudorapidity ($\eta$) and transverse momentum ($p_T$) 
distribution of Higgs + high-$p_T$ jet production. 

In order to obtain the best LO prediction, we 
employ a somewhat different scheme than usual, which particularly
affects the bottom-quark processes.
Instead of discarding higher order contributions 
in the cross section predictions according 
to mere power counting of QCD and electroweak couplings ($g_S,e$), 
regardless of the appearing type of Yukawa coupling ($y_b$, $y_t$),
we argue that keeping the LO contributions to 
each monomial in $y_b$ and  $y_t$
and discarding the higher order contributions to each of them
separately, gives a more accurate approximation to the full result. 
This approach is particularly well-suited 
for the bottom-quark processes, because in this case tree-level
graphs $\propto g_S y_b$ and one-loop graphs $\propto g_S^3 y_t$
contribute and just discarding the loop-contributions
because of the higher power of $g_S$ does not do justice 
to the hierarchy of Yukawa couplings 
$ y_t\gg y_b$
in the 
Standard Model.

As the Higgs plus $b$-jet final state is also of separate interest,
in particular in models beyond the SM, where the top and bottom Yukawa
couplings can be dramatically different, we discuss 
this process also separately in appropriate places below.

Section~\ref{sect:Partonic Processes} 
briefly reviews the contributing parton processes and
Section~\ref{sect:Calculational Approach} 
gives details on our calculational approach. 
In Section~\ref{sect:Numerical results for hadronic cross sections} 
we present numerical results for the distributions
at the LHC and the Tevatron 
and Section~\ref{sect:Conclusions} 
contains our conclusions.

\section{Partonic Processes}
\label{sect:Partonic Processes}

\begin{figure}[tb]
\centerline{
\includegraphics{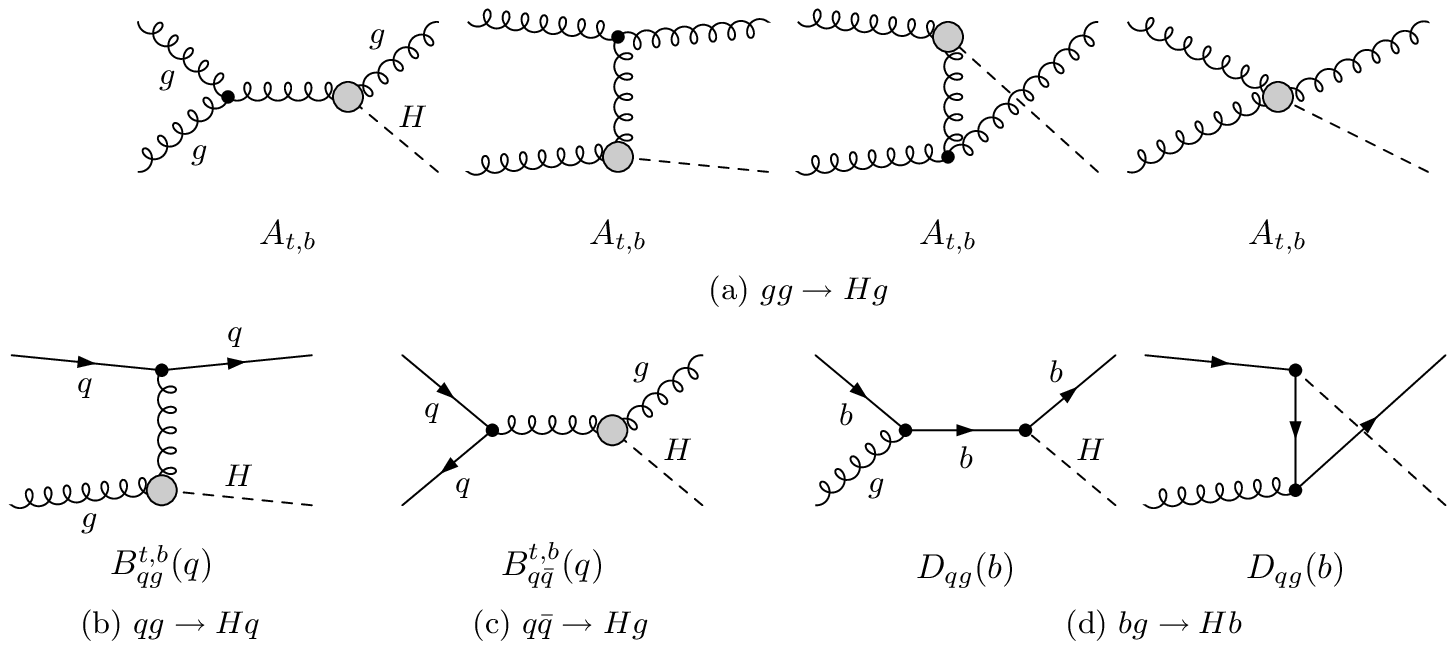}
}

\caption{\label{hjet-QCD} 
QCD contributions to the scattering amplitudes of the 
partonic subprocesses 
(a) $gg \to Hg$, 
(b) $qg \to Hq$, 
(c) $q\bar q\to Hg$ ($q = u,d,s,c$)
and
(d) $bg\to Hb$ at leading order.
The shaded blob represents a quark loop (only top- and 
bottom loops contribute significantly).
The symbol below each graph indicates to which coefficient in the 
mathematical expressions for the amplitudes (Eqs.~(\ref{gg-amp}) and
(\ref{q-amp})) this graph contributes. 
}
\end{figure}

At the partonic level, production of a Higgs boson $H$ together
with a jet at hadron colliders
is mediated by three classes of processes (see Fig.~\ref{hjet-QCD}):
\begin{align*}
& \text{gluon fusion} &   g+g & \to g+H\,,\\
& \text{quark--gluon scattering\hspace*{-2cm}} & q(\bar q)+g &\to q(\bar q)+H\,,\\
& \text{quark--anti-quark annihilation\hspace*{-2cm}} & q +\bar q &\to g+ H\,.
\end{align*}

For the gluon fusion channel, the leading order
contribution to the cross section prediction is 
of order $\alpha_S^3 \alpha$ in the strong and electromagnetic 
coupling constants $\alpha_S$ and $\alpha$.
The corresponding Feynman graphs have one-loop
topologies and are displayed in Fig.~\ref{hjet-QCD}(a), where the shaded 
blobs represent triangle- and box-type loops of virtual quarks.
The loop contributions can be further subdivided according to
their scaling with
the quark Yukawa couplings $y_q = \frac{e}{2 s_W}\frac{m_q}{m_W}$, 
or equivalently
$\alpha_q := y_q^2/4\pi$, 
appearing in  the mathematical expression for the amplitude.
Naturally, the dominant cross section contribution 
($\propto \alpha_S^3\alpha_t$) comes from top-quark loops in the SM.
However, as has been demonstrated e.g.\!\! in \cite{keung-petriello}, 
including bottom-quark loops ($\propto \alpha_S^3\alpha_b$)
changes the gluon fusion cross section 
by several percent.
There is no electroweak contribution
to gluon fusion at one-loop order.

\begin{figure}[tb]
\centerline{
\includegraphics{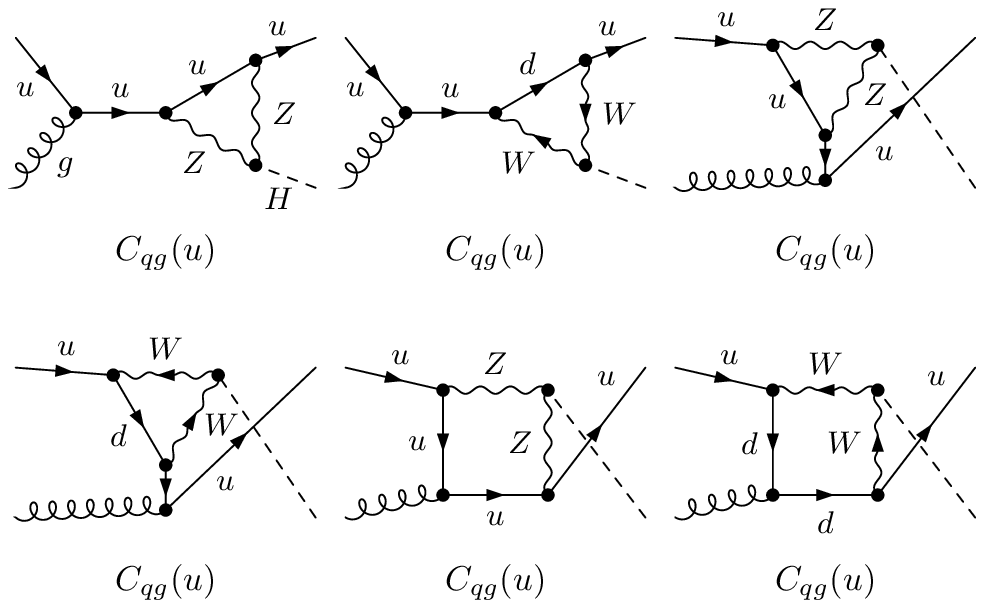}
}

\caption{\label{ughu-EW} 
Electroweak loop contributions to the $u g\to H u$ scattering amplitude
at leading order, assuming no up-quark Higgs Yukawa coupling.
The depicted graphs contribute all to the coefficient $C_{qg}(u)$
in the mathematical expression for the amplitude (Eq.~(\ref{q-amp})).
The contributions look similar for the scattering of 
other quark flavours.
}
\end{figure}

For quark--gluon scattering and quark--anti-quark annihilation
of light quarks ($u,d,s,c$), the leading order QCD cross section
prediction is also loop-induced and of order $\alpha_S^3 \alpha_q$,
where only the $\alpha_t$ and $\alpha_b$ terms are relevant.
The prediction 
is calculated from
the graph topologies displayed in Figs.~\ref{hjet-QCD}(b)
and~\ref{hjet-QCD}(c)
with triangle-type loops of heavy quarks as subgraphs.
The Feynman graphs for the process $\bar qg \to H\bar q$ 
is obtained by crossing the appropriate quark lines 
in Fig.~\ref{hjet-QCD}(b).
In contrast to gluon fusion, these two reactions are also 
mediated by electroweak one-loop graphs $\propto g_S e^3$.
Fig.~\ref{ughu-EW}
exemplarily shows the full set of Feynman graphs 
for the scattering of an up-type quark with a gluon. 
The corresponding Feynman graphs
for the processes $\bar u g\to H \bar u$ and $u \bar u\to H g$ 
can be obtained by crossing the appropriate external lines in
Fig.~\ref{ughu-EW}.

\begin{figure}[tb]
\centerline{
\includegraphics{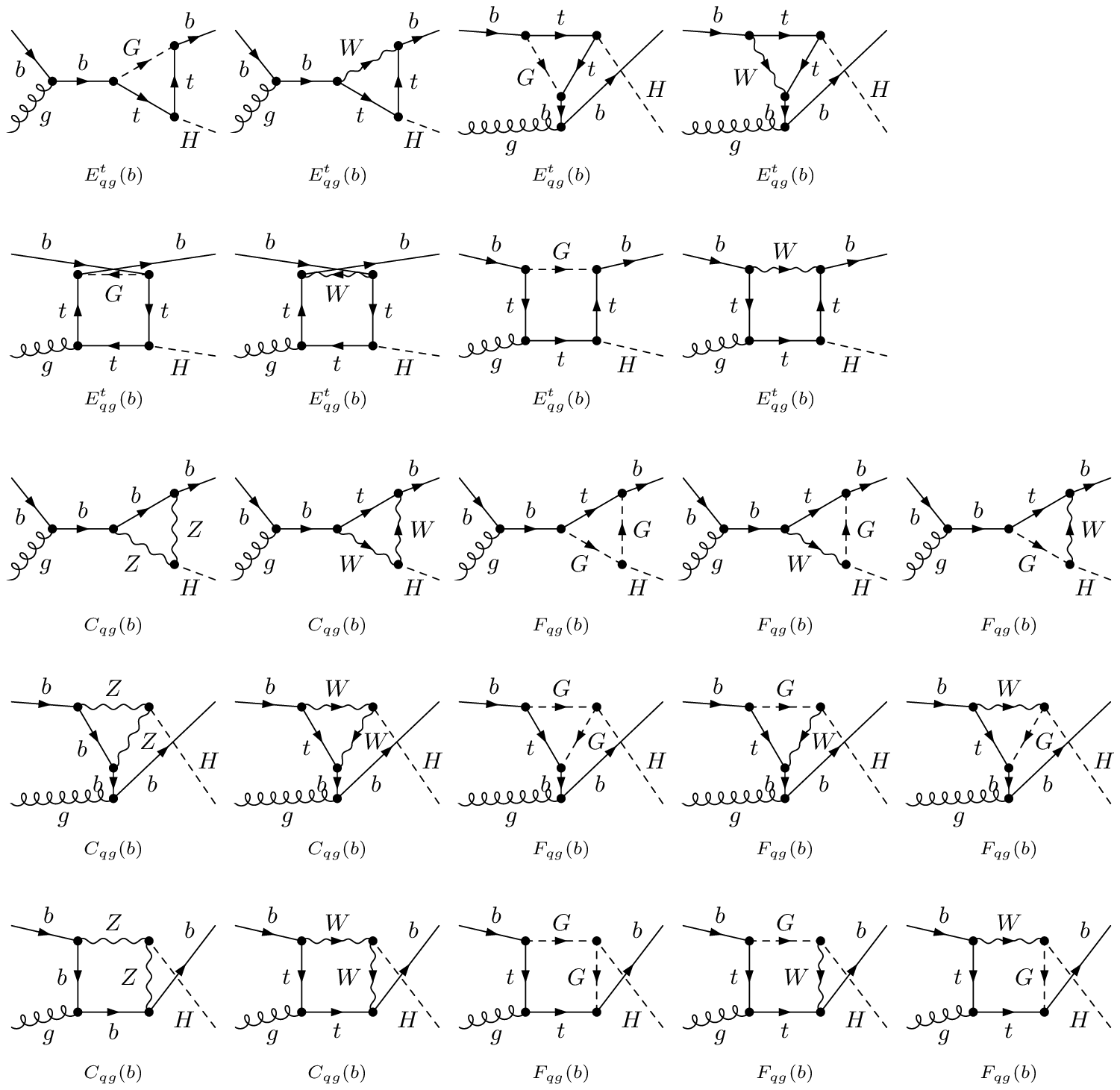}
}

\caption{\label{bghb-EW} 
Electroweak loop contributions to the $b g\to H b$ scattering amplitude
at leading order which do not vanish for $m_b = 0$.
The symbol below each graph indicates to which coefficient in the 
mathematical expression for the amplitude (Eq.~(\ref{b-amp})) 
this graph contributes. 
}
\end{figure}

In the five-flavour scheme, also
$b$-quark initiated
processes for quark--gluon scattering and quark--anti-quark annihilation
contribute to the Higgs plus jet cross section.
For bottom--gluon scattering 
and $b\bar b$ annihilation, 
the QCD leading order cross section prediction 
is $\propto\alpha_S \alpha_b$ and
comes from tree-level Feynman graphs shown in Fig.~\ref{hjet-QCD}(d).
Because of the hierarchy of Yukawa couplings, loop-induced
QCD and electroweak 
contributions to the $b$-quark parton processes,
which contain a factor $\alpha_t$, can be as
important as the tree level contribution in the SM \cite{Mrenna-Yuan}.
The QCD contributions to $bg$ scattering and $b\bar b$ annihilation
are shown in Fig.~\ref{hjet-QCD}(b) 
and Fig.~\ref{hjet-QCD}(c), respectively, where 
the external quarks are $b$ quarks and the shaded blob represents 
triangle-type loops of virtual top-quarks.%
\footnote{The same contribution with a virtual
bottom instead of a top quark 
is only
part of the 
subleading QCD contribution $\propto \alpha_S^2 \alpha_b$ to 
the cross section prediction and cannot be 
considered separately. 
While the interference of this particular graph
with the tree-level QCD amplitude in our approach 
($y_b$ non-zero and $m_b = 0$ otherwise, to be explained in the next Section) 
vanishes, there are more one-loop graphs which contribute to QCD correction
of that order.
}
The Feynman graphs are $\propto g_S^3 y_t$.
The electroweak contributions to $bg$ scattering 
which do not vanish for $m_b=0$
are shown in Fig.~\ref{bghb-EW}. 
The contributions to the $\bar b g\to H \bar b$ and $b \bar b\to H g$ 
amplitudes can be obtained by crossing the appropriate external 
lines in Fig.~\ref{bghb-EW}.
The first two rows of Feynman graphs in Fig.~\ref{bghb-EW} 
are proportional to the top-quark Yukawa coupling $y_t$.
We do not consider graphs $\propto y_b$ here, as they only form a part of
the subleading electroweak contribution $\propto \alpha_S \alpha \alpha_b$
to the cross section prediction and cannot be
considered separately.

The $b$-quark initiated processes are usually not considered 
a relevant contribution to the Higgs plus jet final state in 
the Standard Model (SM), but may even become dominant in models beyond 
the SM with a strongly enhanced bottom-Higgs Yukawa coupling.
This is for instance the case in the Minimal Supersymmetric 
Standard Model (MSSM) for the lightest MSSM Higgs boson 
for low values of the $A$-boson mass $m_A$ ($\lesssim 120\,\gev$) 
\cite{hjet-own,hjet-own-distributions,BField-etal,langenegger-etal}.
However, in our calculation, we include 
the $b$-quark parton processes in the Higgs plus jet cross section
calculation in order to assess the size of their contribution
in comparison with taking effects of a non-vanishing bottom-Higgs
Yukawa-coupling into account in the 
loop-induced light-quark parton processes.

\section{Calculational Approach}
\label{sect:Calculational Approach}

For perturbative predictions involving Higgs bosons, 
the Higgs couplings to other particles, 
which are proportional to their mass, 
create further hierarchies within loop-contributions
in addition to the power series in the QCD and the electromagnetic 
coupling constant $\alpha_S$ and $\alpha$.
Therefore, it is useful to judge the importance 
of contributions also according to the hierarchy of Yukawa coupling
constants $\alpha_q = y_q^2/4\pi = \frac{1}{4\pi}m_q^2/v^2$  ($v=246\,\gev$) 
which is induced by the hierarchy of quark masses $m_q$.
For the two heaviest quarks, this hierarchy is even more pronounced 
than the hierarchy between radiative corrections of consecutive perturbative 
QED orders. In fact, for our choice of mass parameters (see below) we get 
$\alpha_t=3.9\TE{-2}$ and $\alpha_b=2.3\TE{-6}$ and
$\alpha_b/\alpha_t \approx (\alpha(0))^2 < (\alpha_S(m_Z))^4$.

In order to provide an overview of the leading order contributions 
to the Higgs plus jet production channels subdivided by the hierarchy 
of Yukawa couplings, we provide symbolic equations for the scattering 
amplitudes and the squared matrix elements which highlight the 
dependence on the coupling constants.
The scattering amplitudes we consider have the form:
\begin{align}
\label{gg-amp}
{\cal M}_{gg}  &= A_t g_S^3 y_t + A_b g_S^3 y_b \,,\\
\label{q-amp}
{\cal M}_{X}(q)  &= B_X^t(q) g_S^3 y_t + B_X^b(q) g_S^3 y_b 
	+ C_X(q) g_S e^3
\,,\\
\label{b-amp}
{\cal M}_{X}(b)  &= D_X(b) g_S y_b + B_X^t(b) g_S^3 y_t
		+ E_X^t(b) g_S e^2 y_t 
		+ \left( C_X(b) + F_X(b) \right)g_S e^3
\,,
\end{align}
with  $X\in\{qg, \bar q g, q\bar q\}$ and  $q=u,d,s,c$.
The squared matrix elements, only keeping non-vanishing contributions
for $m_b = 0$ while retaining $\alpha_b$ non-zero (explanation follows below), 
read:
\begin{align}
|{\cal M}_{gg}|^2/(4\pi)^4  &=\alpha_t |A_t|^2 \alpha_S^3
	+ \alpha_b |A_b|^2 \alpha_S^3
	+\sqrt{\alpha_t \alpha_b}\: 2 \RE[{A_t}^\star A_b] \alpha_S^3
\,,
\end{align}
\begin{align}
\label{q-amp-sq}
|{\cal M}_{X}(q)|^2/(4\pi)^4  &=\alpha_t 
		 |{B_X^t}(q)|^2 \alpha_S^3
		+\alpha_b 
		    |{B_X^b}(q)|^2 \alpha_S^3
\\\nonumber
		&+\sqrt{\alpha_t\alpha_b}
		    \Big\{
			2 \RE\Big[({B_X^t}(q))^\star {B_X^b}(q)\Big] \alpha_S^3
			\Big\}
\\\nonumber
		& 
		+\sqrt{\alpha_t}
		\Big\{
		2 \RE\Big[({B_X^t}(q))^\star {C_X}(q)\Big] 
			\alpha_S^2\alpha\sqrt{\alpha}
		\Big\}\\\nonumber
		& 
		+\sqrt{\alpha_b}
		\Big\{ 
		2 \RE\Big[({B_X^b}(q))^\star {C_X}(q)\Big] 
			\alpha_S^2\alpha\sqrt{\alpha}
		\Big\} \\\nonumber
		& + |{C_X}(q)|^2 \alpha_S\alpha^3 
\,,
\end{align}
\begin{align}
\label{b-amp-sq}
|{\cal M}_{X}(b)|^2/(4\pi)^4 &=  
	\alpha_t\Big\{
		  |{B_X^t}(b)|^2 \alpha_S^3 
		+ 2\RE\Big[ ({B_X^t}(b))^\star {E_X}^t(b)
			\Big] \alpha_S^2\alpha
		+ |{E_X^t}(b)|^2 \alpha_S \alpha^2 
	\Big\} \\\nonumber
	& +\alpha_b \Big\{
		|D_X(b)|^2\alpha_S
	\Big\} (4\pi)^{-2}\\\nonumber
	& 
	+ \sqrt{\alpha_t}
		\Big\{ 2\RE\Big[({B_X^t}(b ))^\star ({C_X}(b) + {F_X}(b)) \Big]
		\alpha_S^2\alpha \sqrt{\alpha} \Big.\\\nonumber
	& \Big. \hspace{1.5cm} + 2\RE\Big[({E_X^t}(b))^\star ({C_X}(b) + {F_X}(b))\Big]
		\alpha_S \alpha^2  \sqrt{\alpha}
		\Big\}\\\nonumber
	& +|{C_X}(b) + {F_X}(b)|^2 
		\alpha_S \alpha^3 
\,.
\end{align}
For the $b$-quark processes, the interference between 
the tree level graphs
(see Fig.~\ref{hjet-QCD}(d))
and one-loop QCD (see Figs.~\ref{hjet-QCD}(b),(c))
and electroweak graphs (see Fig.~\ref{ughu-EW}) 
vanishes for $m_b=0$.

In order to provide an accurate prediction of all QCD and electroweak
leading order contributions to the Higgs plus jet cross section,
respecting the Yukawa hierarchies, we choose the following approach.
In the squared amplitude of each process we take into account 
the full polynomial in $\sqrt{\alpha_t}$ and $\sqrt{\alpha_b}$,
i.e. we do not discard terms according to mere QCD power counting.
Beyond leading order,
this approach would amount to a reordering of the perturbative expansion,
where corrections to each of the appearing monomials in $\sqrt{\alpha_q}$
(mainly $q=t,b$)
has a separate power series in the QCD and electromagnetic coupling constant.
Furthermore, e.g. for processes with external $b$-quarks, additional
positive powers of $m_b$ ($= v y_b$) can appear in squared matrix elements
from interference terms. In such cases, a consistent expansion of the 
massive-quark squared matrix element in $m_b$, regardless whether it 
occurs in the coupling constant $\sqrt{\alpha_b}$ or directly, 
has to be performed.%
\footnote{
For the bottom-gluon scattering
cross section we study here, we see that the contribution from interference
of the 
tree-level and top-loop QCD graphs is of order $\alpha_b\sqrt{\alpha_t}$
and thus further ``Yukawa-suppressed'' than all other terms we take into
account. This is the formal justification for retaining bottom-mass
dependence only in the Yukawa coupling and using $m_b=0$ otherwise,
which makes the interference contribution vanish.
}

\begin{figure}[ht]
\centerline{
\includegraphics{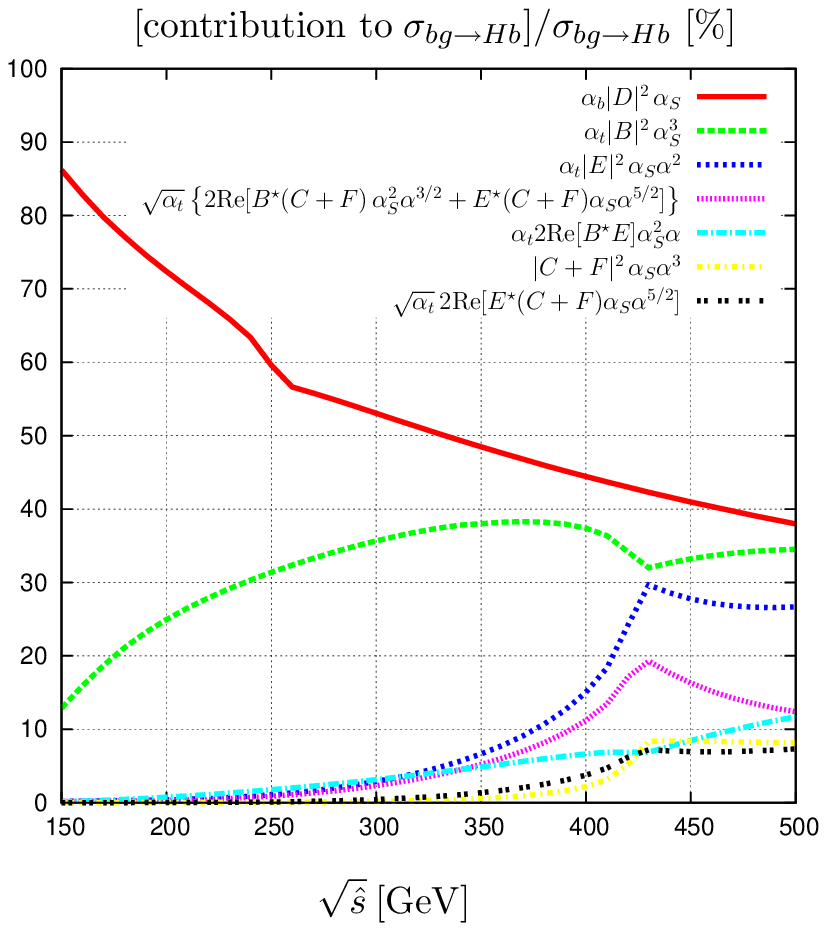}
}
    \caption{
	\label{fig:MEsquared-relative-contributions}
	Relative contributions of different parts 
	of the squared matrix element $|{\cal M}_{qg}(b)|^2$ 
	(see Eq.~(\ref{b-amp-sq}))
	to the integrated partonic cross section of $bg \to Hb$ 
	as a function of centre-of-mass energy $\sqrt{\hat s}$
	for $m_H = 120\,\gev$ and a cut on the scattering angle
	$10^\circ < \hat\theta < 170^\circ$.
        }
\end{figure}

That this approach is adequate for a study of all relevant
leading contributions is confirmed
by the comparably large relative size of the squared and interference 
contributions of the different terms of the amplitude
in Eqs.~(\ref{q-amp-sq}) and (\ref{b-amp-sq}) to the 
cross section prediction.
As an example, Fig.~\ref{fig:MEsquared-relative-contributions}
demonstrates this behaviour for bottom-gluon parton
scattering for $m_H = 120\,\gev$ and a cut on the scattering angle
$10^\circ < \hat\theta < 170^\circ$.
While the top-loop squared contribution is quite large
throughout the displayed centre-of-mass energy $\sqrt{\hat s}$,
also the contribution of the squared electroweak loops reach more than 
25\% for $\sqrt{\hat s} > 425\,\gev$, 
which corresponds to $p_T > 195\,\gev$.
In this range, all other subleading terms still contribute more than 7\%.
For $\sqrt{\hat s} > 330\,\gev$ (i.e. $p_T > 143\,\gev$),
the contribution of the tree-level squared amplitude alone falls 
below 50\%.

In order to represent effects of $b$ quarks as complete as possible,
we always take them into account in quark loop graphs contributing
to gluon fusion and all light quark parton processes 
using the pole mass $m_b$.
Furthermore, we adopt the five-flavour scheme for the description 
of $b$ quark parton process.
Working in the five-flavour scheme, we consider a set of parton distribution
functions (PDFs) which uses a perturbatively defined $b$-quark PDF
where, in the procedure of fitting free parameters in the light-quark and
gluon PDFs to experimental results, also $b$-sensitive observables
have been considered.
Virtually all modern PDF sets are five-flavour PDF sets.
Using $b$-quark PDFs in the five-flavour scheme for predicting the 
$H$ + 1 $b$-jet production cross section is well justified by the 
good
agreement of the NLO QCD result in this scheme 
with the corresponding NLO QCD result
calculated in a four-flavour scheme \cite{bghb-QCD2,LesHouches03-05}.
In the latter approach, among the quark flavours only $u,d,s,c$ are
considered to be distributed in the proton while final state 
bottom quarks can only appear through gluon-initiated processes.

We consider the $b$-quark to be a massless parton to be consistent
with the parton model but retain a non-zero Yukawa coupling $y_b$.
Specifically for the $b$ quark parton process,
we use the running $\overline{MS}$ mass at NLO QCD
for the bottom quark,
$$
  m_b^2(\mu_R) = m_b^2 
\left\{1-2\,\frac{\alpha_S}{\pi}\left[
	\log\left(\frac{\mu_R^2}{m_b^2}\right) + \frac{4}{3}
	\right]\right\}
\,,
$$
as the mass parameter in $y_b$ and we choose the factorisation 
scale $\mu_F^{(b)}=m_h/4$. In this way, the prediction for the 
tree level bottom-gluon scattering cross section 
gives a good approximation to the NLO QCD result
\cite{bghb-QCD2,DSSW,bscale-choice}.

\section{Numerical results for hadronic cross sections}
\label{sect:Numerical results for hadronic cross sections}

Hadron colliders like the LHC (Tevatron) collide protons with protons
(anti-protons) 
with a total energy $\sqrt s$
in the laboratory frame.
Hadronic cross sections are obtained via convolution of the
parton-level cross sections with the parton distribution functions (PDFs) 
and summation over the various contributing partons.
Experimental restrictions to the detectability
of the produced particles are conventionally realized by
imposing specific cuts to the kinematically allowed phase space.
Typically, cuts are imposed on the final-state
transverse momentum and/or the pseudorapidity in order 
to have high-$p_T$ jets not too close to the beam axis.
In our case, we choose the selection criteria,
\begin{align}
\label{the-cuts}
p_T & > p_T^\MIN\,, & 
| \eta_{\JET} | & < \eta_{\text{max}}\,,
\end{align}
where $p_T$ and $\eta_{\JET}$
denote transverse momentum  and pseudorapidity
of the final state parton.
We evaluate the differential hadronic cross sections
$d\sigma_{AB}/d p_T$ and $d\sigma_{AB}/d \eta_{\JET}$ (with $AB=pp$ or
$p\bar p$) 
in the presence of those cuts as described in detail in
\cite{hjet-own-distributions}.

The numerical evaluation has been carried out with the NLO MSTW2008 
PDFs~\cite{MSTW2008}
and a consistently chosen strong coupling constant $\alpha_S(\mu_R)$,
i.e. using the formula 
including the NLO QCD corrections (see e.g. Ref.~\cite{pdg2006})
for $n_f=5$ with $\alpha_S(m_Z)=0.1201789$.
The renormalisation
scale $\mu_R$ and factorisation scale for the gluon and the light quarks
$\mu_F$ are chosen both equal to $m_h$. 
For the bottom-quark factorisation scale $\mu_F^{(b)}$
we choose $m_h/4$.
In the calculation, the pole masses for the top and bottom quark
are set to the values
\begin{align*}
m_t&= 171.3\,\gev\,, & m_b&=4.2\,\gev\,,
\end{align*}
while the values of the electroweak parameters 
are chosen as 
\begin{align*}
m_W&= 80.398\,\gev\,, & m_Z&=91.1876\,\gev\,, 
	& G_F&= 1.16637\TE{-5}\,\gev^{-2}\,,\\
&& \cos\theta_w & = \frac{m_W}{m_Z}\,, 
	& \alpha & = \frac{\sqrt 2}{\pi}G_F m_W^2 \sin^2\theta_w\,.
\end{align*}
The CKM matrix is assumed to be diagonal.
We have evaluated the partonic cross sections
with the help of the computer programs 
FeynArts 3.2 and FormCalc 3.2\footnote{The results are 
corrected for a documented bug (see file ``ChangeLog'' in FormCalc version
5.3 or newer for details) in the calculation of a particular 
type of colour factor appearing in the gluon-fusion amplitude.} \cite{FAFC} and further 
convoluted them with PDFs according to formulae 
given in \cite{hjet-own-distributions}.
For the   evaluation of the NLO MSTW2008 PDFs~\cite{MSTW2008} 
and the corresponding $\alpha_S(\mu_R)$ we used LHAPDF \cite{lhapdf}.

\subsection{Tevatron}
\label{sect:Tevatron}

For the numerical evaluation of all cross sections 
we use $\sqrt{s}=1.96\,\tev$ and
the cuts in Eq.~(\ref{the-cuts})
with $p_T^\MIN = 15\,\gev$ (as in Ref.~\cite{keung-petriello})
and $\eta_{\text{max}} = 2.5$, i.e. we use the former (latter)
in the calculation of $\eta$ ($p_T$) distributions.
At the Tevatron, the contributions of quark-gluon scattering
and gluon fusion to the Higgs plus jet cross section
are comparable in size, e.g. for $m_H=120\,\gev$ and the given cuts,
quark-gluon scattering contributes in total 41.5\% 
and bottom-gluon scattering alone 1.4\%. 
Furthermore, $q\bar q$ annihilation contributes in total 4.5\%
and $b\bar b$ annihilation alone 0.23\%.
We start by presenting the impact of electroweak loops
and $b$-quark processes on the differential cross section predictions
for quark-gluon scattering and quark--anti-quark annihilation separately.

\subsubsection{Quark-Gluon Scattering}
\label{sect:Quark-Gluon Scattering}

\begin{figure}[ht]
\includegraphics{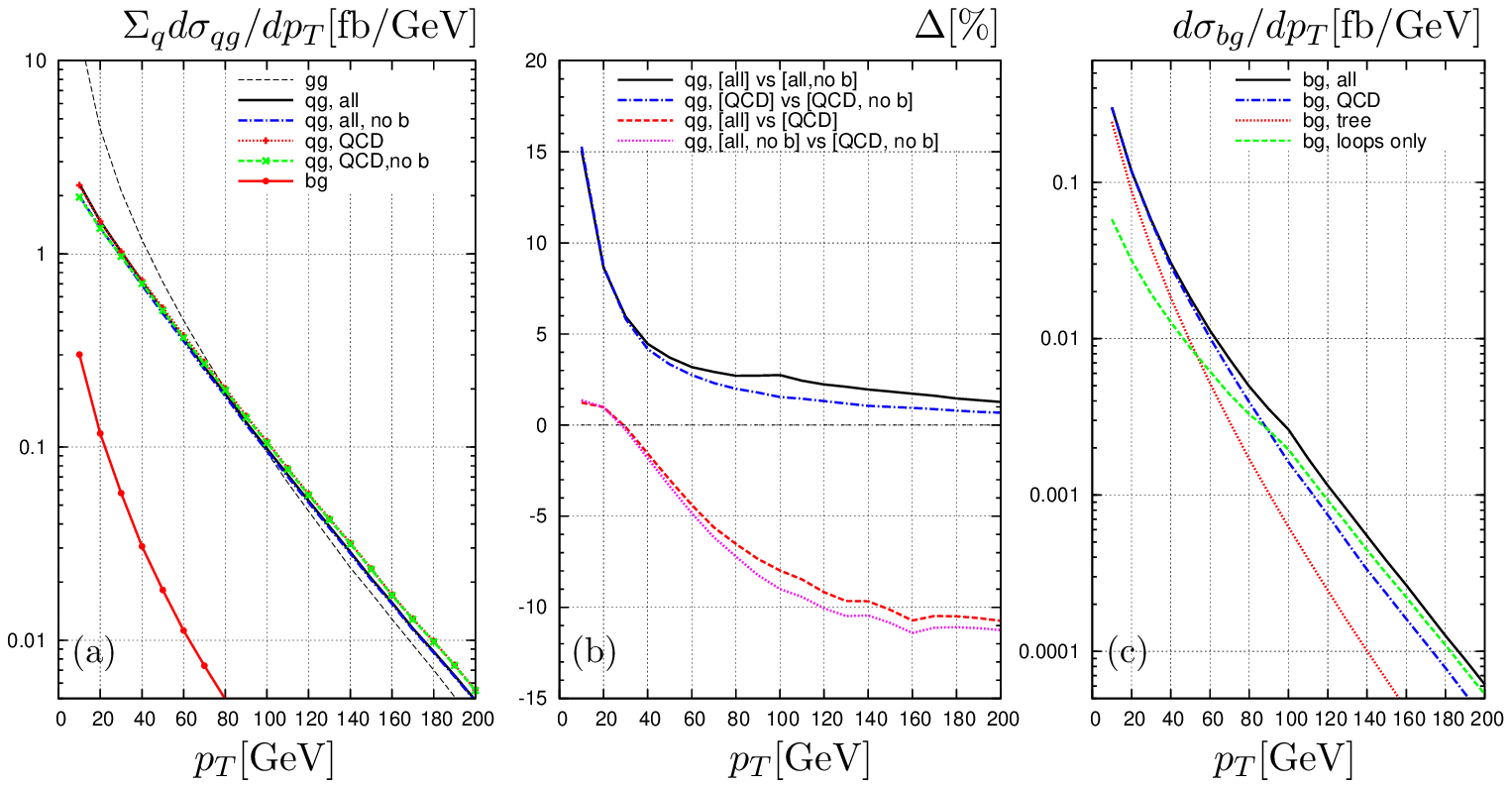}
    \caption{
	\label{fig:pt.mh120.qg.tev}
        $p_T$ distribution for quark-gluon 
	scattering at the Tevatron: 
	(a) quark parton processes with and without the $b$ quark
	contributions,  
	    with and without electroweak contributions;
	(b) relative differences to the left panel;
	(c) contributions to the $b$ quark parton processes.
	The depicted approximations are described in the main text.
        }
\end{figure}

Fig.~\ref{fig:pt.mh120.qg.tev}(a) shows the jet
transverse momentum ($p_T$) distribution of Higgs plus jet production via
quark-gluon scattering calculated in four approximations: 
either including or neglecting electroweak loop contributions
combined with
either including or neglecting the $b$-quark parton contributions
in the prediction, represented by the identifiers 
``{\tt all}'', ``{\tt all, no b}'', ``{\tt QCD}'' and ``{\tt QCD, no b}'' 
in the Figure.
For comparison, the contributions of gluon fusion (including top- and
bottom-loop graphs in the amplitude) and of bottom-gluon scattering
to the Higgs plus jet cross section
are also shown in Fig.~\ref{fig:pt.mh120.qg.tev}(a).
Among the parton contributions to the Higgs plus jet $p_T$ distribution,
quark--gluon scattering
is the largest contribution 
for $p_T$  above about $80\,\gev$, while gluon fusion dominates 
at lower $p_T$.
The relative contribution of electroweak loops is very small 
at low $p_T$ and increases with rising $p_T$, while it is 
the other way around with the contributions of $b$-quark
parton processes.

Fig.~\ref{fig:pt.mh120.qg.tev}(b)
shows 
relative differences $\Delta$ between the four
approximations to the quark-gluon scattering cross section
displayed in Fig.~\ref{fig:pt.mh120.qg.tev}(a).
The effect of including electroweak contributions depends on $p_T$
(dashed line) and the overall effect ranges from +5\% to -12\% 
for $10\,\gev < p_T <200\,\gev$.
The effect of the electroweak loops contributing to the $b$-quark processes 
amounts to a slight rise of the Higgs plus jet cross section by at most 
2\%.
It is approximately given 
by the difference between the dashed and dotted lines in
Fig.~\ref{fig:pt.mh120.qg.tev}(b), where the latter represents 
the relative effect of electroweak loops in the light-quark parton processes 
only which corresponds to (and agrees with) 
results of Ref.~\cite{keung-petriello}.

Fig.~\ref{fig:pt.mh120.qg.tev}(c) shows 
the 
bottom-gluon scattering cross section alone calculated in four 
approximations.
While the identifier ``{\tt bg, all}'' in Fig.~\ref{fig:pt.mh120.qg.tev}(c)
corresponds to 
retaining all contributions to the amplitude non-zero in Eq.~(\ref{b-amp}),
``{\tt bg, QCD}'' corresponds to setting $E,C,F$
to zero, i.e. the amplitude is approximated by the tree-level graphs
in Fig.~\ref{hjet-QCD}(d) and the top-loop graphs in
Fig.~\ref{hjet-QCD}(b).
The identifier ``{\tt bg, tree}'' indicates that only the tree-level amplitude
is considered, i.e. only $D$ in Eq.~(\ref{b-amp}) is non-zero,
and ``{\tt bg, loops only}'' corresponds to setting only $D$ to zero.
In the $p_T$ range displayed, none of the contributions to
bottom-gluon scattering is negligible.

For the prediction of the $p_T$ distribution of $bg$ scattering, 
it turns out that discarding the squared top-loop graphs 
in the cross section prediction ($\propto\alpha_t$ in Eq.~(\ref{b-amp-sq})),
because their contribution is formally of higher order in $\alpha_S$, 
as has often been done in the past,
is a bad approximation for larger $p_T$ values.
Although this contribution to the partonic $bg$ cross section is smaller 
than the tree-level contribution at low centre-of-mass energy 
$\sqrt{\hat s}$, it has a flatter energy dependence.
Thus, it becomes larger than the tree-level contribution beyond a certain
energy $\sqrt{\hat s_0}$.
For instance, for $m_H = 120\,\gev$ and a cut on the scattering angle
$10^\circ < \hat\theta < 170^\circ$,
we get $\sqrt{\hat s_0} = 430\,\gev$
and for $\sqrt{\hat s} > 260\,\gev$ the top-loop contribution 
is already more than $50\%$ 
of the tree-level contribution 
(see Fig.~\ref{fig:MEsquared-relative-contributions}).
The situation is similar for the contribution of electroweak loop graphs.
Taking those into account as well, the turnover point where the 
loop-squared becomes larger than the tree-squared is
$\sqrt{\hat s_0} = 330\,\gev$ and for $\sqrt{\hat s} > 230\,\gev$ 
the loop contribution is $> 50\%$ of the tree-level contribution.
As final state jets with a certain $p_T$ introduce the energy threshold
\begin{align}
\label{jet-energy-threshold}
\sqrt{\hat s} > p_T + \sqrt{m_H^2+p_T^2} =: \sqrt{\hat s_{\text{low}}}
\,,
\end{align}
the above behaviour implies that for $p_T > 198\, [143]\,\gev$ 
top-quark [electroweak and top-quark] loop contributions dominate 
the $p_T$ distribution.
Therefore, we argue that these loop contributions, which do not 
interfere with the tree-level cross section for $m_b=0$, need to 
be taken into account for an accurate leading order prediction
of the bottom--gluon scattering contribution to Higgs plus jet.
While a mere QCD power counting would suggest otherwise, clearly,
the hierarchy of Yukawa couplings in the SM also suggests not to
discard loop-squared parts $\propto \alpha_t$ in the cross section
prediction.
\medskip

\begin{figure}[hbt]
\includegraphics{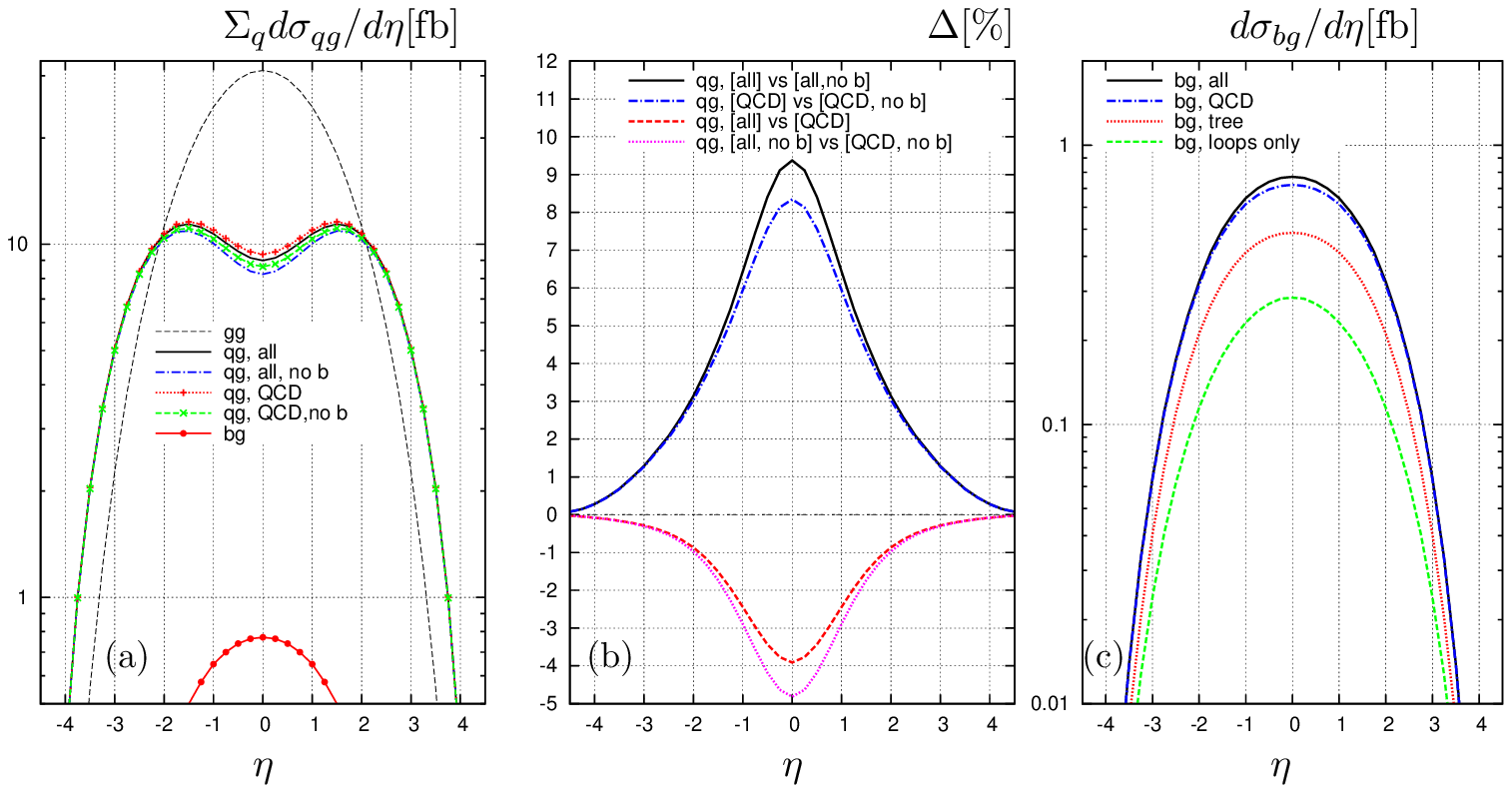}
    \caption{
	\label{fig:eta.mh120.qg.tev}
        $\eta$ distribution for quark-gluon 
	scattering at the Tevatron: 
	(a) quark parton processes with and without the $b$ quark 
        contributions,
	    with and without electroweak contributions;
	(b) relative differences to the left panel;
	(c) contributions to the $b$ quark parton processes.
	The depicted approximations are described in the main text.
        }
\end{figure}

Fig.~\ref{fig:eta.mh120.qg.tev} shows the same breakdown of contributions
as the previous Figure, now for the pseudorapidity distribution 
$d\sigma/d\eta$ of the final state jet.
Electroweak and $b$-quark parton contributions
depend strongly on $\eta$ 
(dashed and solid line in Fig.~\ref{fig:eta.mh120.qg.tev}(b))
and are roughly of the same order
(but opposite sign) with extrema of about $-4\%$
and 9\%, respectively, at $\eta = 0$, i.e. 
for a jet radiated into the central part of the detector. 
The explanation of the strong $\eta$ dependence of these contributions
has two elements. 
One element is the appearance of Feynman graph topologies 
in the amplitudes with different angular dependence
than in the dominant light-quark QCD amplitudes. 
The other is the different angular dependence in the laboratory frame
caused by parton collisions which have preferentially boosted final states
due to unbalanced valence quark contributions 
(double-peak in $d\sigma/d\eta$ for light-quark--gluon scattering) 
as compared to parton collisions with no valence quarks involved 
(central peak in $d\sigma/d\eta$ for bottom--gluon scattering).

The $bg$ scattering contribution to the $\eta$ distribution 
in Fig.~\ref{fig:eta.mh120.qg.tev}(c) illustrates how
much the large contribution of loop-squared terms (QCD top-loop and 
electroweak) to the $p_T$ distribution in Fig.~\ref{fig:pt.mh120.qg.tev}(c) 
contribute significantly to $d\sigma/d\eta$, i.e. after $p_T$ integration 
for given $\eta$.
The bottom-gluon scattering cross section prediction
including electroweak and top-loop graphs in the amplitude (solid line)
is about 40\% larger than the tree-level prediction (dotted line) at $\eta =0$. 
The bulk of the contributions comes from including the top-loop 
graphs (dot-dashed line) which leads to a prediction that
deviates from the full result for $bg$ scattering
only by about $-5\%$.

\subsubsection{Quark--Anti-Quark Annihilation}
\label{sect:Tevatron Quark--Anti-Quark Annihilation}
\begin{figure}[hbt]
\includegraphics{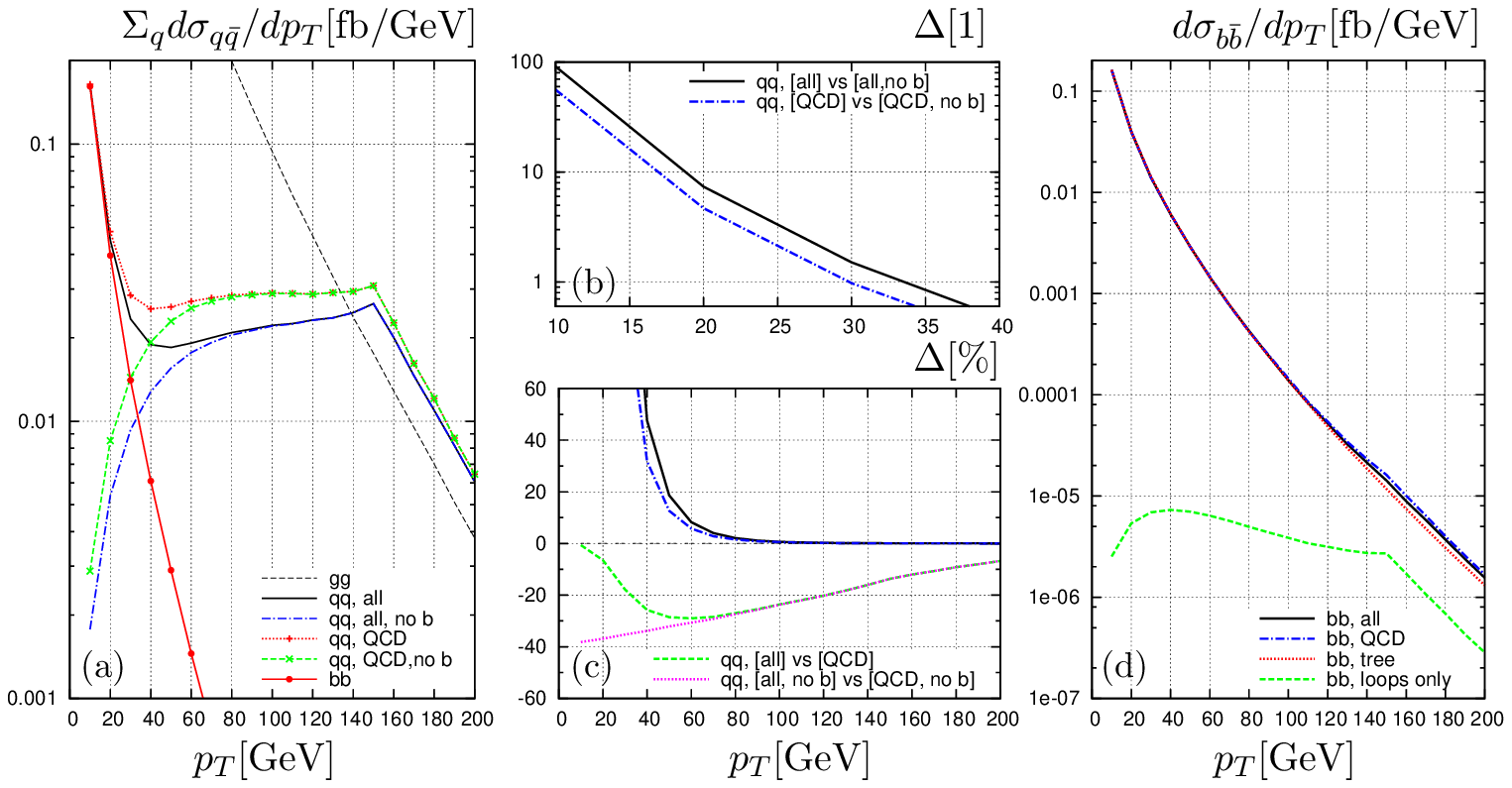}
    \caption{
	\label{fig:pt.mh120.qq.tev}
        $p_T$ distribution for 
	quark--anti-quark annihilation
	at the Tevatron: 
        (a) quark parton processes with and without the $b$ quark
        contributions,
            with and without electroweak contributions;
	(b),(c) relative differences to the left panel;
	(d) contributions to the $b$ quark parton processes.
	The depicted approximations are described in the main text.
        }
\end{figure}

Fig.~\ref{fig:pt.mh120.qq.tev}(a) shows the jet $p_T$
distribution via the $q\bar q$ annihilation channel
calculated in the same four approximations as described
in the previous Section (\ref{sect:Quark-Gluon Scattering}). 
Also here, the hadronic cross section via gluon fusion 
and $b\bar b$ annihilation are shown for comparison.
Bottom quarks dominate the $q\bar q$ contribution 
to Higgs+jet
below $p_T \approx 35\,\gev$.
The reason for this dominance is that 
$b\bar b$ annihilation
is mediated at tree-level by $t$- and $u$-channel virtual $b$-quark 
exchange graphs (see Fig.~\ref{hjet-QCD}(d)) 
which make the cross section large at small non-zero 
$p_T$ 
and it would rise indefinitely for $p_T \to 0$.
In contrast to that, the contributions from light-quark parton processes 
vanish in this limit.
This rise of the $b\bar b$ annihilation cross section 
for small $p_T$ is analogous to the situation for the 
gluon-fusion and quark-gluon scattering process, 
where $t$- or $u$-channel virtual gluon exchange 
graphs cause a similar behaviour. In the low $p_T$ region,
soft gluon resummation is called for in order to obtain a 
reliable prediction.
However, we do not expect that the relative sizes of different
contributions to the Higgs plus jet cross section, 
which is our main concern here, are altered 
greatly by resummation effects.

The contribution of $q\bar q$ annihilation reaches 
the size of gluon fusion above $p_T \approx 140\,\gev$.
Remarkably, this contribution
rises with $p_T$ up to about $150\,\gev$ despite the fact that the 
parton luminosity is falling at the same time.
This behaviour is caused by a pronounced threshold peak 
in the energy dependence of 
the partonic cross sections for $q\bar q\to Hg$ ($q=u,d,s,c$) 
which enters the phase space integration over jet 
final states with a given $p_T$.
The threshold peak is located around $\sqrt{\hat s} = 2 m_t
=: \sqrt{\hat s_{\text{peak}}}$
and is due to the top-loop graphs
of the QCD part of the amplitude \cite{BaurGlover}.
The influence of this peak on the hadronic $p_T$ distributions
is strong if it occurs near the energy threshold
for jets with given $p_T$: $\sqrt{\hat s_{\text{low}}}$ in
Eq.~(\ref{jet-energy-threshold}).
Equating $\sqrt{\hat s_{\text{peak}}} = \sqrt{\hat s_{\text{low}}}$,
we get the transverse momentum where the peak exactly occurs
at the boundary of the integration over jet phase space: 
$$
p_T^{\text{peak}}=m_t\left(1-\frac{m_H^2}{4m_t^2}\right)\approx 150\,\gev\,.
$$
Approaching $p_T^{\text{peak}}$ from below,
the double effect of the PDFs falling 
with rising $\sqrt{\hat s_{\text{low}}}$
and the cross section for 
$\sqrt{\hat s} \gtrsim \sqrt{\hat s_{\text{low}}}$
rising because of the peak moving closer towards $\sqrt{\hat s_{\text{low}}}$
is an overall rise of the $p_T$ distribution.
For $p_T > p_T^{\text{peak}}$, the peak region falls outside of the 
region of jet phase space integration and thus $d\sigma/dp_T$ drops sharply.

Fig.~\ref{fig:pt.mh120.qq.tev}(b) and \ref{fig:pt.mh120.qq.tev}(c) show
relative differences between results of the four approximations. 
In particular, 
Fig.~\ref{fig:pt.mh120.qq.tev}(b) zooms in on the special area $p_T <
40\,\gev$. Naturally, as the quark annihilation
contribution of $b$-quarks rises and of 
light quarks vanishes with $p_T\to 0$, respectively, the 
relative contribution of the $b$-quark processes (solid line)
reaches almost a factor of 100 at $p_T=10\,\gev$.
While the electroweak contributions to the light-quark processes
lead to large and negative interference (dotted line) at low $p_T$
(e.g. $-37\%$ for $p_T=20\,\gev$), 
the electroweak contributions to
the $b$-quark processes 
are much smaller in the same $p_T$ region (e.g. $+2\%$ for $p_T=20\,\gev$).
Hence, with increasing dominance of the $b$-quark processes
towards lower $p_T$, 
the relative contribution of electroweak loops in the sum
of all quarks (dashed line) is diminished.
We confirm the results for the electroweak loop effects in 
the light-quark $q\bar q$ annihilation processes \cite{keung-petriello} 
but also demonstrate the importance of including the $b$-quark
initiated processes as well.

In Fig.~\ref{fig:pt.mh120.qq.tev}(d),
a breakdown of different contributions to the $p_T$ distribution of the
$b\bar b$ annihilation process 
analogous to the one in Fig.~\ref{fig:pt.mh120.qg.tev}(c),
described in Section \ref{sect:Quark-Gluon Scattering},
demonstrates 
the smallness of all loop contributions below $p_T\approx 120\,\gev$.
From this information, one can already conclude 
that the $\eta$ distribution for the given cut of $p_T > 15\,\gev$ 
will be practically independent of the loop contributions.
The hadronic cross section via $b\bar b$ using only loop graphs 
in the amplitude (dashed line),
shows qualitatively similar behaviour with $p_T$ as the light-quark
annihilation cross section. This is also due to a  
threshold peak caused by top-loop graphs
in the partonic scattering amplitude, as explained above.
\medskip

\begin{figure}[hbt]
\includegraphics{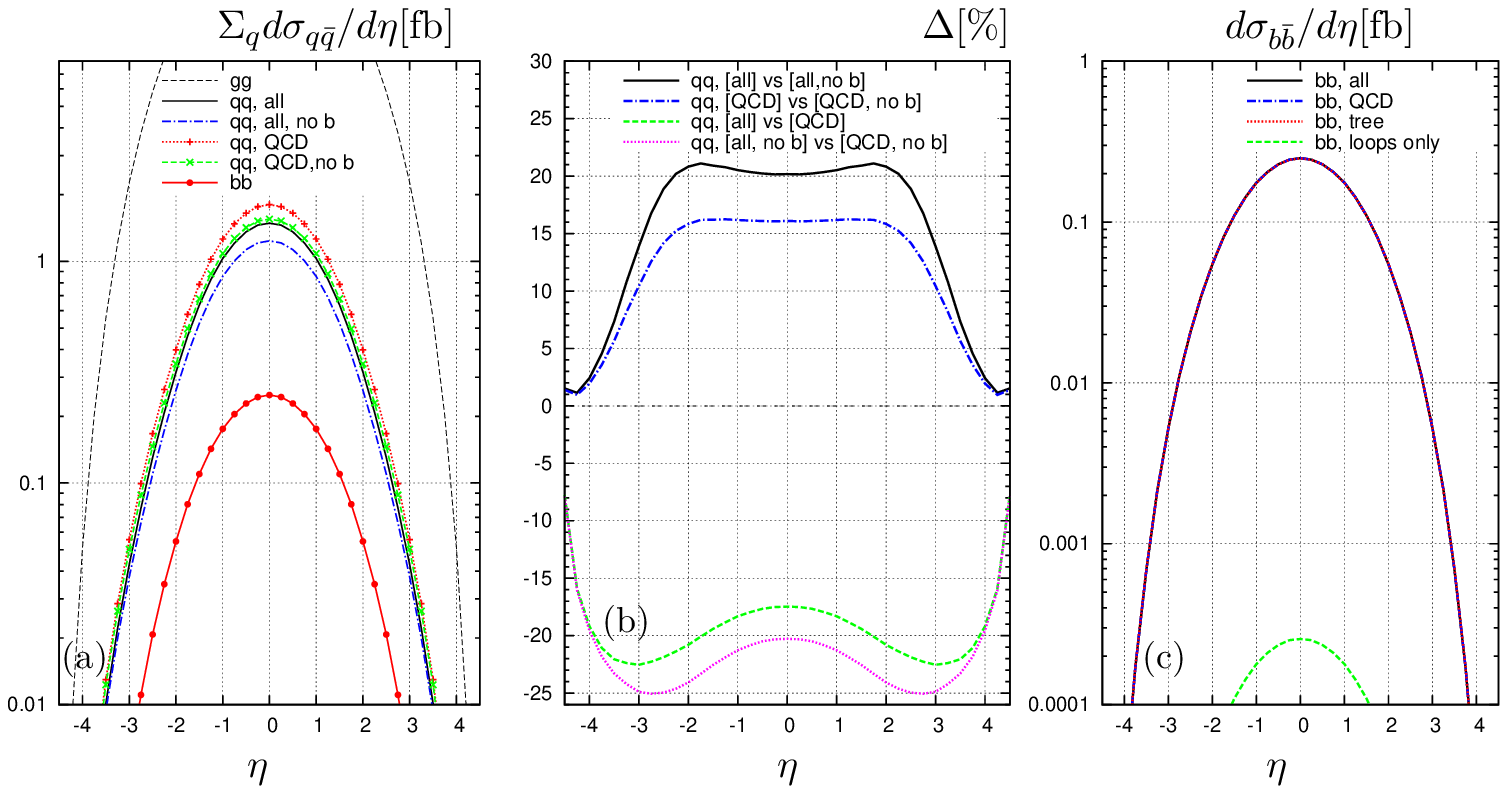}
    \caption{
	\label{fig:eta.mh120.qq.tev}
        $\eta$ distribution for 
	quark--anti-quark annihilation
	at the Tevatron: 
        (a) quark parton processes with and without the $b$ quark
        contributions,
	    with and without electroweak contributions;
	(b) relative differences to the left panel;
	(c) contributions to the $b$ quark parton processes.
	The depicted approximations are described in the main text.
        }
\end{figure}

In Fig.~\ref{fig:eta.mh120.qq.tev},
the $\eta$ distribution of the $q\bar q$ annihilation channel is shown
split in separate contributions as in previous figures.
The hadronic cross section via $q\bar q$ calculated taking electroweak
and $b$-quark contributions into account (solid line in
Fig.~\ref{fig:eta.mh120.qq.tev}(a)) and neglecting them both
($\times$-dashed line) are roughly equal.
This shows that for the given kinematical cuts, electroweak and  $b$-quark
contributions roughly compensate each other, as can be seen more 
quantitatively in Fig.~\ref{fig:eta.mh120.qq.tev}(b).
Fig.~\ref{fig:eta.mh120.qq.tev}(c) demonstrates the expected 
smallness of the loop contributions to the $b\bar b$ channel.

\subsubsection{Corrections to Higgs + Jet Distributions}
\begin{figure}[hbt]
\includegraphics{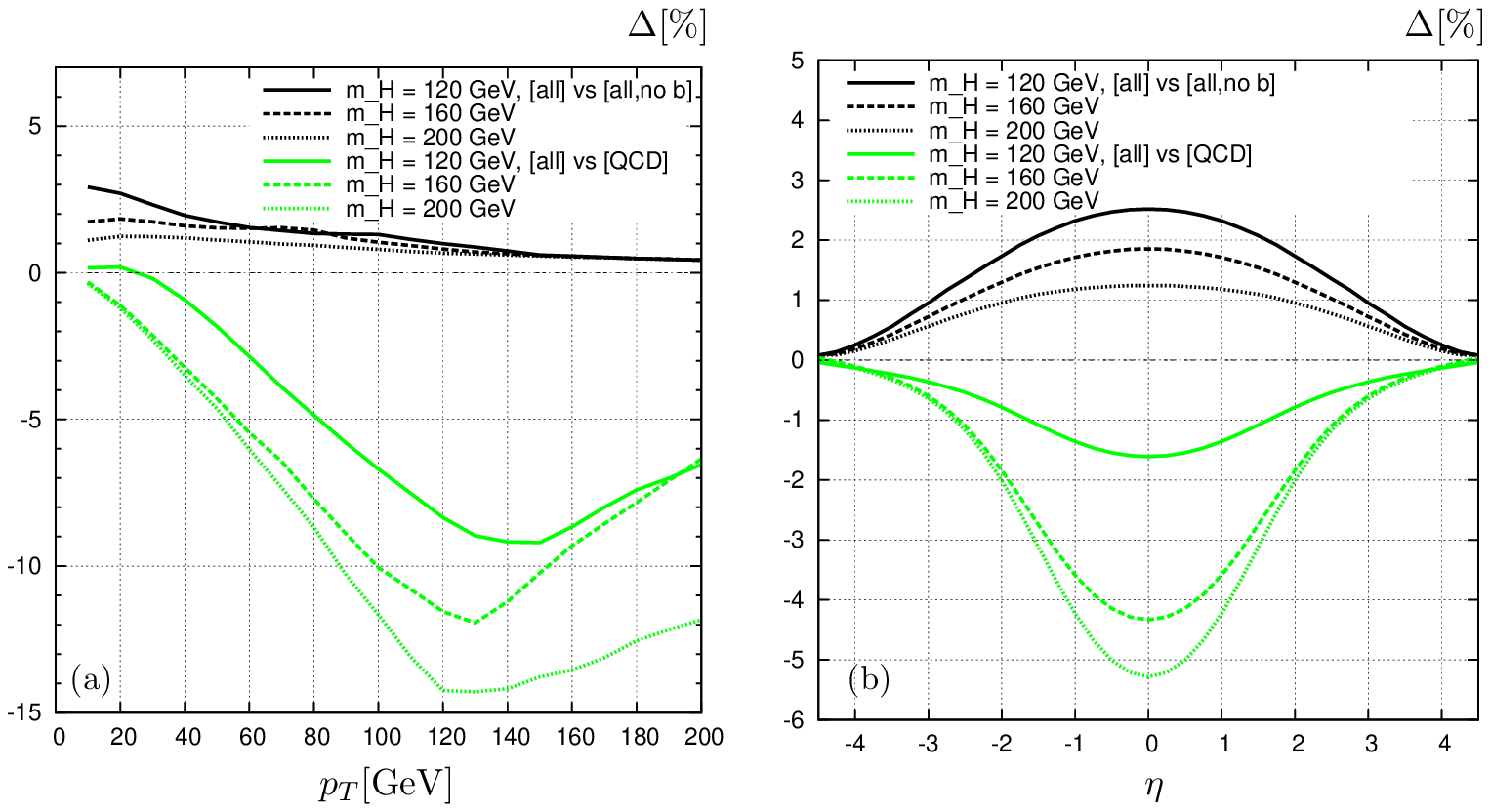}
    \caption{
	\label{fig:reldiff.tev}
	Relative size of the effects of including $b$-quark parton processes
	(black lines) and electroweak contributions (green lines) on the 
	Tevatron ($\sqrt{s} = 1.96\,\tev$) differential distributions for 
        Higgs + Jet production: (a) $p_T$ distribution
	and (b) $\eta$ distribution of the recoiling jet.
        }
\end{figure}

We discuss the impact of contributions from electroweak loops and 
$b$-quark parton processes on the total 
Higgs plus high-$p_T$ jet cross section,
i.e. the sum of the gluon fusion, quark-gluon scattering
and $q\bar q$ annihilation contribution.
Figs.~\ref{fig:reldiff.tev}(a) and \ref{fig:reldiff.tev}(b) 
show the relative size of the contributions to the $p_T$ and
$\eta$ distribution, respectively,
for three different 
Higgs masses: $m_H = 120\,\gev$, $160\,\gev$ and $200\,\gev$.
A feature, both present in the  $p_T$ and the $\eta$ distribution, 
is that the $b$-quark contributions decrease and 
electroweak contributions increase with rising Higgs mass.
In general, effects on the $p_T$ distribution 
(see Fig.~\ref{fig:reldiff.tev}(a))
by $b$-quark contributions (black lines) decrease 
and electroweak contributions (green lines) increase with $p_T$.
The largest effect of $b$-quark processes 
amounts to an increase of about 3\% at low $p_T$ for $m_H = 120\,\gev$. 
The largest effect of the electroweak contributions 
is a decrease of about $-14\%$ for $p_T$-values in 
the range 110 to $160\,\gev$ and $m_H = 200\,\gev$.
\medskip 

Fig.~\ref{fig:reldiff.tev}(b) shows the impact of electroweak and 
$b$-quark contributions on the $\eta$ distribution.
Both contributions have their largest magnitude at $\eta = 0$:
$+2.5\%$ for $m_H=120\,\gev$ for the $b$-quark contribution
and 
$-5.3\%$ for $m_H=200\,\gev$ for the electroweak contribution.

\subsection{LHC}

For the numerical evaluation of all cross sections
we use $\sqrt{s}=10\,\tev$ and
the cuts in Eq.~(\ref{the-cuts})
with $p_T^\MIN = 30\,\gev$ 
and $\eta_{\text{max}} = 4.5$
which have been used in previous SM
studies for the LHC~\cite{ADIKSS,Zmushko}.
At the end of Section \ref{sect:LHC:Corrections to Higgs + Jet Distributions} 
we make some comments on the $\sqrt{s}$ dependence of the results.
At the LHC, gluon fusion dominates largely over 
quark-gluon scattering and the $q\bar q$ annihilation contribution
is quite small,
e.g. for $m_H=120\,\gev$ and the given cuts, quark-gluon scattering 
and $q\bar q$ annihilation contribute 27.8\% and 0.8\%, respectively,
to the total Higgs plus jet cross section.
The contribution of $bg$scattering and $b\bar b$ annihilation alone
is 2\% and 0.3\%, respectively.
Note that detailed descriptions of certain common features of the 
results which have been already given for the Tevatron case in
Section \ref{sect:Tevatron} will not be given for the LHC case.

\subsubsection{Quark-Gluon Scattering}

\begin{figure}[hbt]
\includegraphics{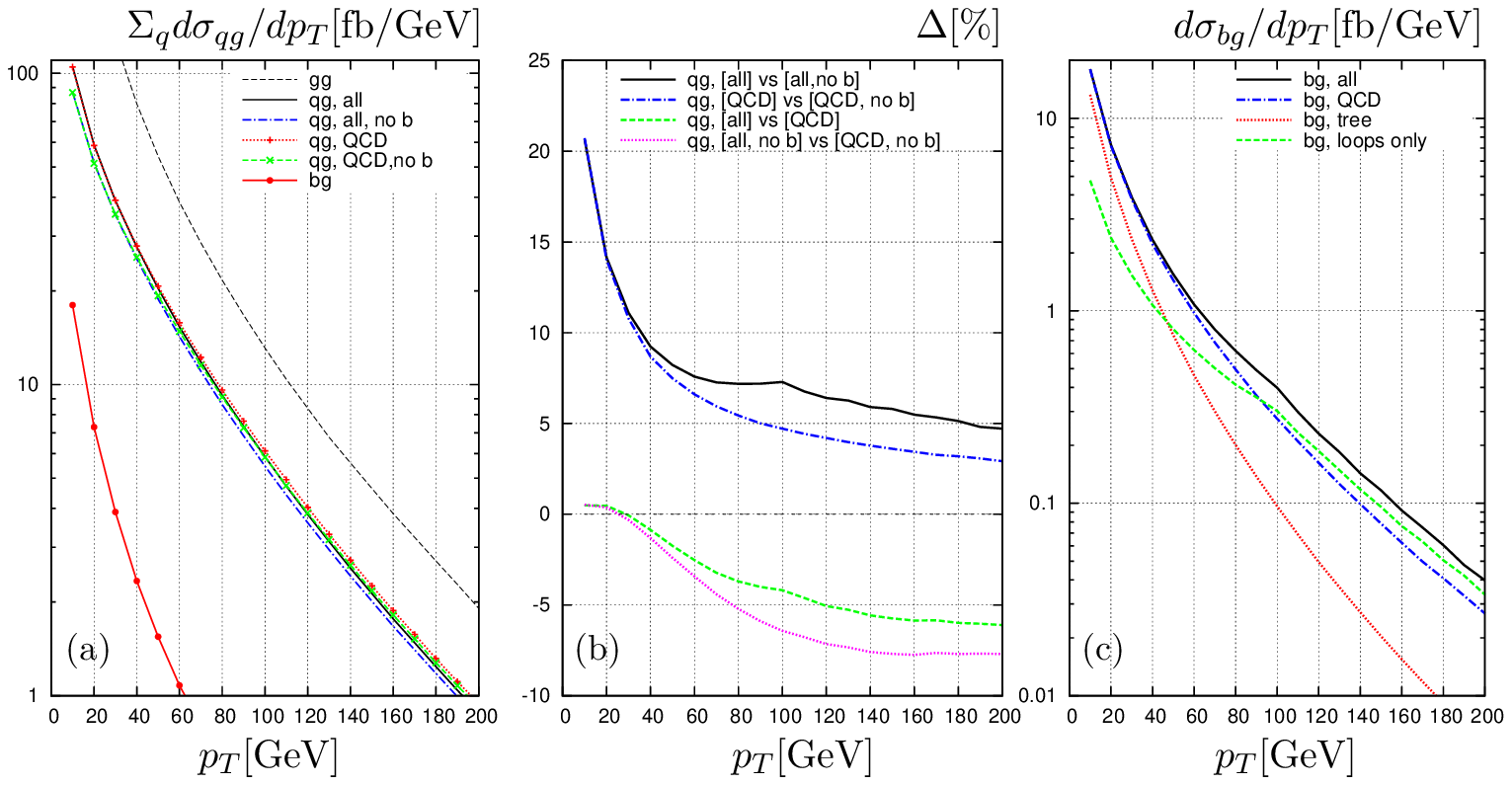}
    \caption{
	\label{fig:pt.mh120.qg.LHC}
        $p_T$ distribution for quark-gluon 
	scattering at the LHC ($\sqrt{s} = 10\,\tev$):
        (a) quark parton processes with and without the $b$ quark
        contributions,
	    with and without electroweak contributions;
	(b) relative differences to the left panel;
	(c) contributions to the $b$ quark parton processes.
	The depicted approximations are described in the main text.
        }
\end{figure}

Fig.~\ref{fig:pt.mh120.qg.LHC}(a) shows the 
$p_T$ distribution of Higgs plus jet production via
quark-gluon scattering calculated in the four approximations 
(``{\tt all}'', ``{\tt all, no b}'', ``{\tt QCD}'' and ``{\tt QCD, no b}'') 
described in Section \ref{sect:Quark-Gluon Scattering}.
The hadronic cross section via gluon fusion 
and bottom-gluon scattering
are also shown in Fig.~\ref{fig:pt.mh120.qg.LHC}(a) for comparison.

Similar to the Tevatron case, the relative contribution 
of electroweak loops, shown in Fig.~\ref{fig:pt.mh120.qg.LHC}(b),
is very small at low $p_T$ and increases with rising $p_T$, 
while it is the other way around for the $b$-quark contributions.
The electroweak contribution (dashed line) ranges
from $+0.4\%$ to $-6\%$ in the displayed 
$p_T$ interval from 10 GeV to 200 GeV.
The relative $b$-quark contribution 
(solid line) ranges from 20\% at low to 4.7\% at high $p_T$.

The effect of the electroweak loops contributing to the $b$-quark processes,
which cause an increase of about $40\%$ of the cross section for these 
subprocesses for $p_T > 100\,\gev$ (see Fig.~\ref{fig:pt.mh120.qg.LHC}(c)),
amounts to a slight rise of the total Higgs+jet cross section by at most 2.5\%
for $p_T$ above $100\,\gev$. 

Fig.~\ref{fig:pt.mh120.qg.LHC}(c) shows the 
bottom-gluon scattering cross section alone calculated in four 
approximations 
(``{\tt bg, all}'', ``{\tt bg, QCD}'', ``{\tt bg, tree}'', 
and ``{\tt bg, loops only}'')
described in Section \ref{sect:Quark-Gluon Scattering}.
In the $p_T$ range displayed, similar to the Tevatron case, 
none of the contributions is negligible,
confirming results of Ref.~\cite{Mrenna-Yuan}.
\medskip

\begin{figure}[hbt]
\includegraphics{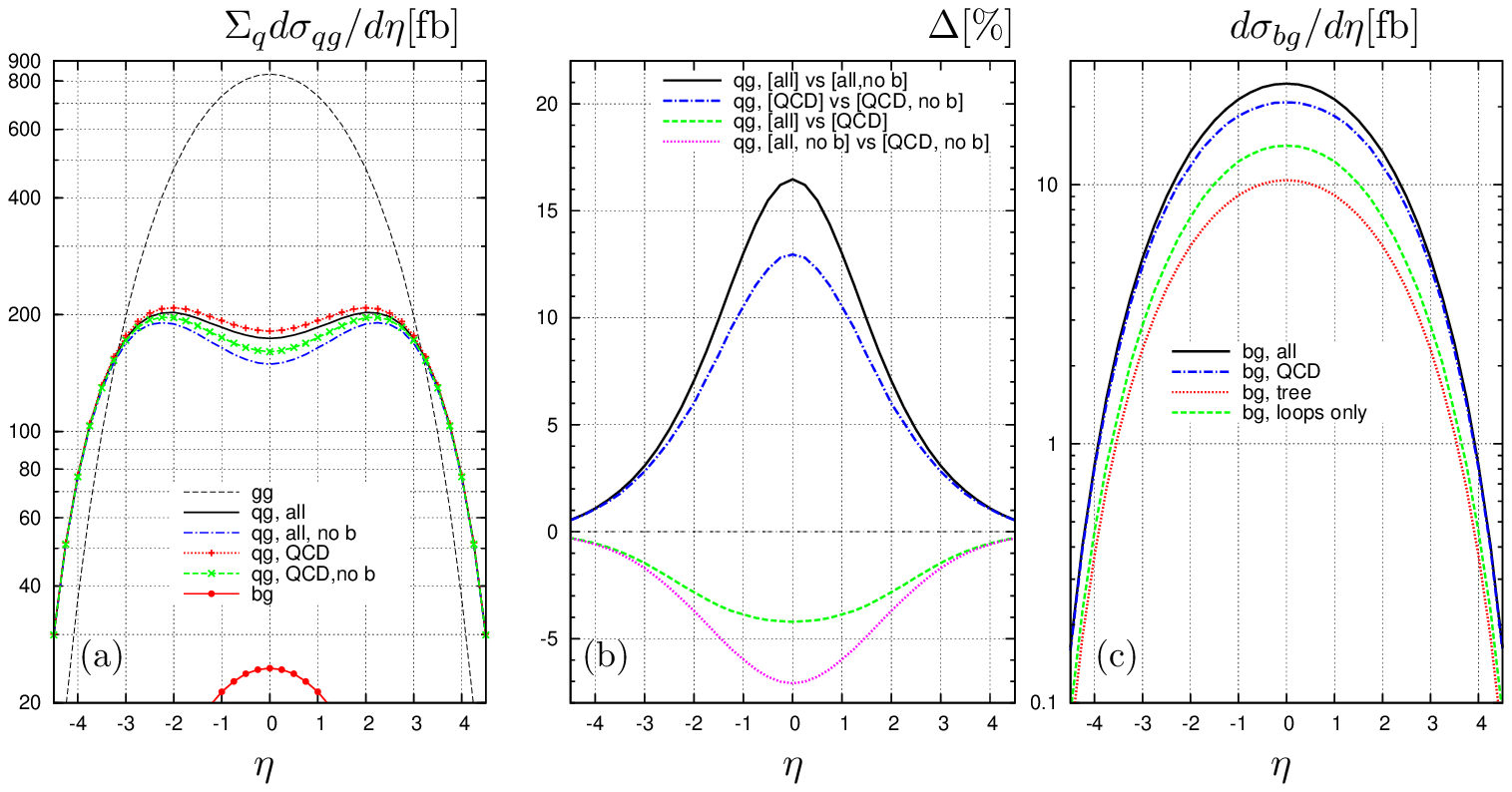}
    \caption{
	\label{fig:eta.mh120.qg.LHC}
        $\eta$ distribution for 
	quark-gluon scattering 
	at the LHC ($\sqrt{s} = 10\,\tev$):
        (a) quark parton processes with and without the $b$ quark
        contributions,
	    with and without electroweak contributions;
	(b) relative differences to the left panel;
	(c) contributions to the $b$ quark parton processes.
	The depicted approximations are described in the main text.
        }
\end{figure}

Fig.~\ref{fig:eta.mh120.qg.LHC} shows the same breakdown of contributions
as the previous Figure, for the jet pseudorapidity distribution 
$d\sigma/d\eta$.
Electroweak and $b$-quark parton contributions
depend strongly on $\eta$ 
(dashed and solid line in Fig.~\ref{fig:eta.mh120.qg.LHC}(b))
and 
have extrema of about $-4\%$
and $+16.5\%$, respectively, at $\eta = 0$.

\subsubsection{Quark--Anti-Quark Annihilation}

\begin{figure}[hbt]
\includegraphics{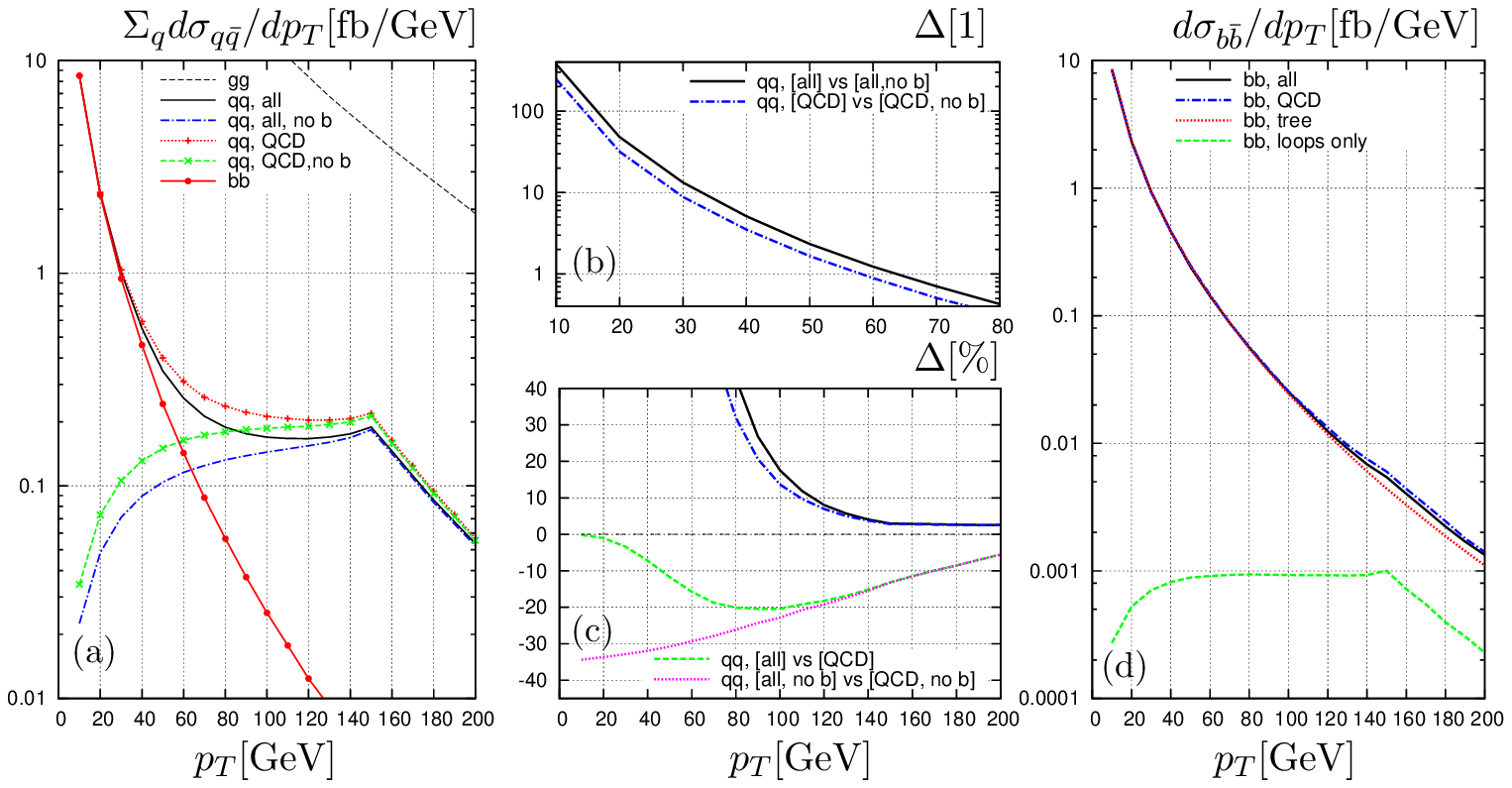}
    \caption{
	\label{fig:pt.mh120.qq.LHC}
        $p_T$ distribution for 
	quark--anti-quark annihilation
	at the LHC ($\sqrt{s} = 10\,\tev$):
        (a) quark parton processes with and without the $b$ quark
        contributions,
	    with and without electroweak contributions;
	(b) relative differences to the left panel;
	(c) contributions to the $b$ quark parton processes.
	The depicted approximations are described in the main text.
        }
\end{figure}

Fig.~\ref{fig:pt.mh120.qq.LHC}(a) shows the jet $p_T$
distribution via the $q\bar q$ annihilation channel
calculated in the same four approximations as described
in Section \ref{sect:Quark-Gluon Scattering}
and of the gluon fusion
and $b\bar b$ annihilation channel for comparison.
As the qualitative behaviour of this channel is similar 
to the Tevatron case, we refer 
to Section~\ref{sect:Tevatron Quark--Anti-Quark Annihilation} 
for their explanation.
The $b\bar b$ contribution 
dominates the $q\bar q$ contribution to the Higgs+jet cross section 
below $p_T \approx 65\,\gev$.
This dominance is 
even more pronounced than in the Tevatron
prediction (see Section \ref{sect:Tevatron Quark--Anti-Quark Annihilation})
and is again due to tree-level $t$- and $u$-channel virtual $b$-quark
exchange graphs (see Fig.~\ref{hjet-QCD}(d)).
This contribution would rise indefinitely
for $p_T \to 0$ while the contribution from light-quark parton processes
vanish in this limit.

Fig.~\ref{fig:pt.mh120.qq.LHC}(b) and \ref{fig:pt.mh120.qq.LHC}(c) show
relative differences between results of the four approximations. 
In particular, 
Fig.~\ref{fig:pt.mh120.qq.LHC}(b) zooms in on the special area $p_T <
80\,\gev$. 
Naturally, as the annihilation contribution of $b$-quarks rises
and of light quarks vanishes with $p_T\to 0$, respectively, the 
relative contribution of the $b$-quark processes (solid line)
reaches a factor of about 370 at $p_T=10\,\gev$.
However, the $b$-quark contribution is not only relevant 
in the low $p_T$ region, in which soft gluon resummation is 
called for in order to obtain a reliable prediction.
It stays above 20\% up to $p_T\approx 95\,\gev$ which is where 
it first falls behind the electroweak contributions
in significance.

While the electroweak contributions to the light-quark processes
lead to large and negative interference (dotted line) at low $p_T$
(e.g. $-34\%$ for $p_T=20\,\gev$), 
confirming results of \cite{keung-petriello}, 
the electroweak contributions to
the $b$-quark processes alone, 
in the same $p_T$ region, are very small (e.g. $+1\%$ for $p_T=20\,\gev$).
Hence, the increasing dominance of the $b$-quark processes
towards lower $p_T$
leads to a decrease of the relative contribution of electroweak loops 
in the sum of all quark parton contributions (dashed line).

In Fig.~\ref{fig:pt.mh120.qq.LHC}(d),
a breakdown of different contributions to the $p_T$ distribution of the
$b\bar b$ annihilation process 
analogous to Fig.~\ref{fig:pt.mh120.qg.LHC}(c) demonstrates 
that sizeable loop contributions to this process
only set in above $p_T\approx 120\,\gev$.
\medskip

\begin{figure}[hbt]
\includegraphics{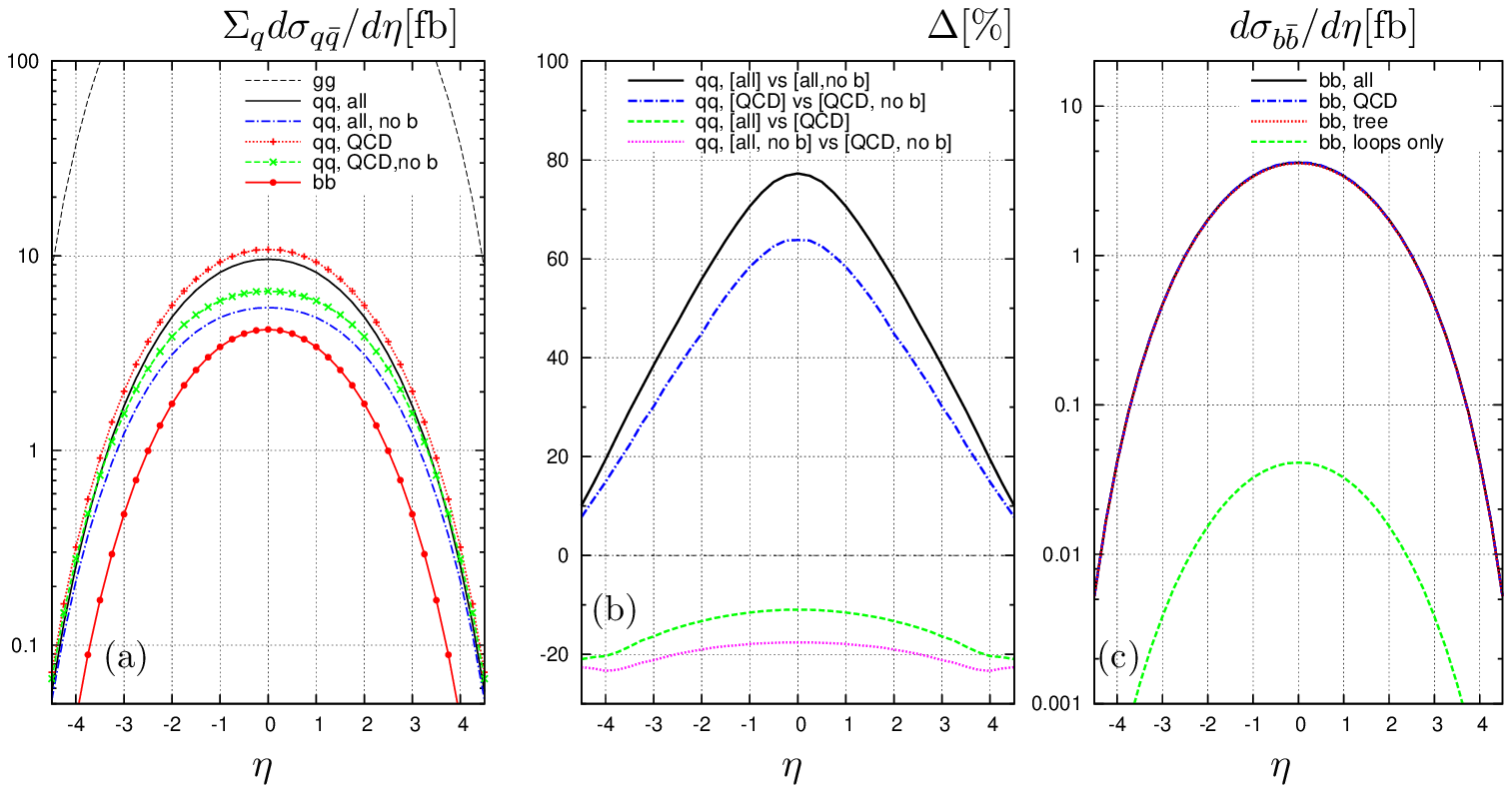}
    \caption{
	\label{fig:eta.mh120.qq.LHC}
        $\eta$ distribution for 
	quark--anti-quark annihilation
	at the LHC ($\sqrt{s} = 10\,\tev$):
        (a) quark parton processes with and without the $b$ quark
        contributions,
	    with and without electroweak contributions;
	(b) relative differences to the left panel;
	(c) contributions to the $b$ quark parton processes.
	The depicted approximations are described in the main text.
        }
\end{figure}

In Fig.~\ref{fig:eta.mh120.qq.LHC},
the $\eta$ distribution of the $q\bar q$ annihilation channel is shown
split in separate contributions as in previous Figures.
The fact that electroweak and $b$-quark contributions are sizeable
is reflected by the clear separation of all lines in 
Fig.~\ref{fig:eta.mh120.qq.LHC}(a).
For the given cut of $p_T > 30\,\gev$, the $b$-quark contribution
(solid line in Fig.~\ref{fig:eta.mh120.qq.LHC}(b))
still reaches 77\% at $\eta = 0$, while the electroweak contribution
(dashed line) varies between $-20\%$ and $-11\%$ 
with a minimum at $\eta = 0$.
Similar to the Tevatron case, loop contributions do not 
affect the $\eta$ distribution of the $b\bar b$ channel
for the given cuts (see Fig.~\ref{fig:eta.mh120.qq.LHC}(c)).

\subsubsection{Corrections to Higgs + Jet Distributions}
\label{sect:LHC:Corrections to Higgs + Jet Distributions}

\begin{figure}[hbt]
\includegraphics{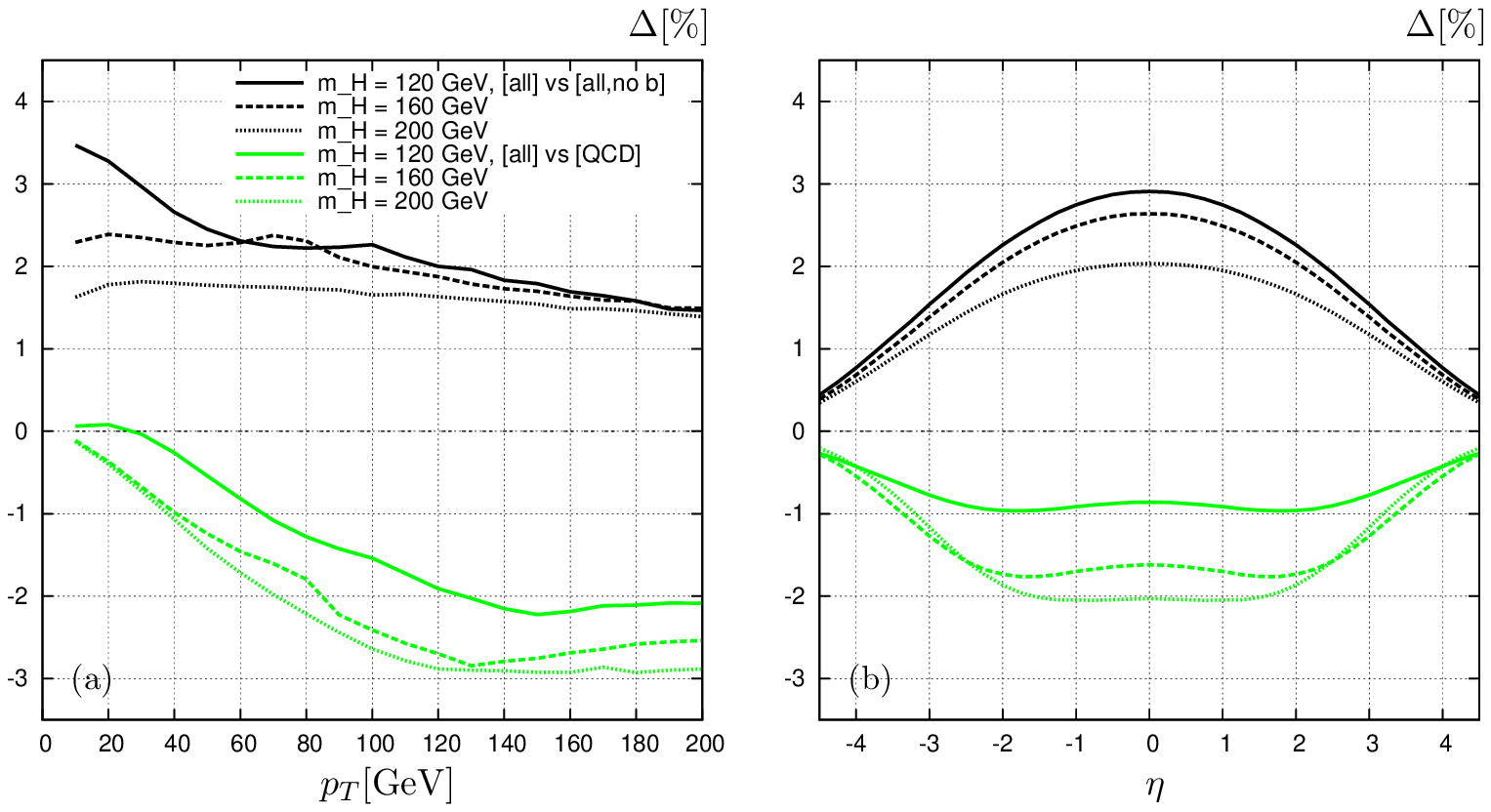}
    \caption{
	\label{fig:reldiff.LHC}
	Relative size of the effects of including $b$-quark parton processes
	(black lines) and electroweak contributions (green lines) on the 
	LHC ($\sqrt{s} = 10\,\tev$) differential distributions for 
        Higgs + Jet production: (a) $p_T$ distribution
	and (b) $\eta$ distribution of the recoiling jet.
        }
\end{figure}

Having studied the effects of electroweak loops and 
$b$-quark processes on  quark-gluon scattering
and $q\bar q$ annihilation separately, 
we discuss here the impact of those contributions
on the total Higgs plus high-$p_T$ jet cross section.
Figs.~\ref{fig:reldiff.LHC}(a) and \ref{fig:reldiff.LHC}(b) 
show the relative size of the contributions to the $p_T$ and
$\eta$ distribution, respectively,
for three different 
Higgs masses: $m_H = 120\,\gev$, $160\,\gev$ and $200\,\gev$.
The qualitative features of the $b$-quark and  electroweak 
contributions to the $p_T$ and $\eta$ distributions are 
similar to the Tevatron case, e.g. $b$-quark decrease and electroweak 
contributions  increase with rising Higgs mass, respectively.
On the quantitative level, the maximal effect of the electroweak loops
on the $p_T$ distribution (green lines in Fig.~\ref{fig:reldiff.LHC}(a))
is diminished to a $-3\%$ change of the cross section in the range $p_T > 120\,\gev$ and for $m_H =
200\,\gev$, compared to $-14\%$ for the Tevatron case.
The reason for this difference is that 
the $qg$ and $q\bar q$ channels, which are affected
at leading order  by electroweak loops, dominate the Higgs + Jet cross section
for large $p_T$ at the Tevatron, while at the LHC (with $\sqrt{s}=10\,\tev$)
gluon fusion dominates over the full $p_T$ range.
Contrary to that, the $b$-quark contributions to the 
$p_T$ distribution (black lines)
are  slightly higher throughout the displayed 
$p_T$ range, reaching maximally $+3.5\%$ at low $p_T$ for $m_H = 120\,\gev$.
\medskip 

Fig.~\ref{fig:reldiff.LHC}(b) shows the impact of electroweak and 
$b$-quark contributions on the $\eta$ distribution.
Both contributions have their largest magnitude at $\eta = 0$:
$+2.9\%$ for $m_H=120\,\gev$ for the $b$-quark contribution
and 
$-2.0\%$ for $m_H=200\,\gev$ for the electroweak contribution.
The large difference between the electroweak contributions
for larger $p_T$ values in the Tevatron and LHC case
is not seen in the $\eta$ distributions.
This is to be expected, as the $p_T$ range where the difference in
$d\sigma/dp_T$ is substantial contributes only little 
to the cross section $d\sigma/d\eta$.
\bigskip

As the collision energy of the LHC will probably be changed 
a few times before the design goal of $14\,\tev$ is reached,
we have also studied the $\sqrt s$ dependence 
of the electroweak and 
$b$-quark contributions to the total Higgs plus jet cross section
(with the given cuts: $p_T^\MIN = 30\,\gev$ and $\eta_{\text{max}} = 4.5$)
for the three Higgs masses $120\,\gev$, $160\,\gev$ 
and $200\,\gev$. We find that the relative contributions 
depend only mildly on the collider energy. 
For instance,
for $\sqrt s$ from $7\,\tev$ to $14\,\tev$, 
the electroweak contributions for $m_H=200\,\gev$ 
vary from -2.2\% to -1.5\%  
and the  $b$-quark contributions for $m_H=120\,\gev$
vary from +2.3\% to +2.45\%.
As is clear from Figs.~\ref{fig:reldiff.LHC}(a) and \ref{fig:reldiff.LHC}(b),
a tighter $p_T$-cut will increase the relative electroweak contribution
while the relative $b$ quark contribution will stay almost the same.
A tighter $\eta$-cut will increase the relative size of both contributions.

\section{Conclusions}
\label{sect:Conclusions}

We have calculated specific contributions to 
jet pseudorapidity and transverse momentum distributions
for the Standard Model Higgs plus high-$p_T$ jet production cross section
at the LHC and the Tevatron.
The remaining scale uncertainty of the NLO QCD prediction
(in the large top-mass limit)
for this Higgs production mode 
via light quark parton processes
is at the level of 10\% \cite{SMHJatNLO,kulesza-etal}.
Motivated by this, we discussed here
the contributions of electroweak loops
and of bottom-quark parton
processes (in the five-flavour scheme) 
to cross section predictions 
for the Tevatron and the Large Hadron
Collider with $10\,\tev$ collision energy. 

For Higgs bosons with a mass of $120\,\gev$, $160\,\gev$ and $200\,\gev$, 
we find the maximal effects 
of the electroweak contributions to the 
total Higgs plus jet $p_T$ and $\eta$ distribution to be 
$-14\%$ and $-5.3\%$, respectively, for the Tevatron, 
and
$-3\%$ and $-2\%$, respectively, for the LHC.
For the maximal contribution of bottom-quark parton processes
to the $p_T$ and $\eta$ distribution, we find
$+3\%$ and $+2.5\%$, respectively, for the Tevatron,
and
$+3.5\%$ and $+3\%$, respectively, for the LHC.
Those contributions are smaller but still comparable in size to
the numerical impact of including bottom quarks of non-zero mass
in quark-loop mediated contributions to the Higgs plus jet
cross section prediction.
For both colliders, the magnitude of the electroweak contribution
rises with Higgs mass, while the bottom parton
contribution falls with Higgs mass.

A separate study of the Higgs + $b$-jet cross section
demonstrates that a calculational approach which respects the 
hierarchies of Yukawa couplings yields a leading order
cross section prediction which is more accurate 
in the high-$p_T$ regime than conventional approaches.

A computer code corresponding to the calculation  
presented here will be made available with the next
version release of the public code HJET \cite{HJET-URL}.

\subsection*{Acknowledgements} 
I thank Frank Petriello for the efficient comparison of our 
numerical results.
I thank Stefan Dittmaier, Alberto Guffanti, Karl Jakobs and 
Christian Schwinn for valuable comments and useful discussions.
This work was supported by the Helmholtz Alliance HA-101 
'Physics at the Terascale' and in part by the 
European Community's Marie-Curie Research
Training Network under contract MRTN-CT-2006-035505
`Tools and Precision Calculations for Physics Discoveries at Colliders'
(HEPTOOLS).

\newpage

\end{document}